\documentclass[showpacs,preprintnumbers,prd,superscriptaddress,nofootinbib, twocolumn]{revtex4}

\usepackage{amsmath}
\usepackage{amsfonts}
\usepackage{amssymb}
\usepackage{graphicx}
\usepackage[titletoc]{appendix}
\usepackage[rightcaption]{sidecap}
\usepackage{subfigure}
\usepackage{soul}

\usepackage{dcolumn}
\usepackage{units}
\usepackage{array}
\usepackage{ctable}
\usepackage{multirow}
\usepackage{siunitx}
\usepackage{longtable}
\usepackage{tabularx}
\usepackage{booktabs}

\graphicspath{{Graphics/}}

\def\be{\begin{equation}}
\def\ee{\end{equation}}
\def\bea{\begin{eqnarray}}
\def\eea{\end{eqnarray}}
\renewcommand{\d}{\mathrm{d}}

\begin{document}



\title{Thermodynamic length, geometric efficiency and Legendre invariance}

\author{Carlo Cafaro}
\email{cafaroc@sunypoly.edu}
\affiliation{SUNY Polytechnic Institute,  257 Fuller Road, 12203 Albany, New York, USA.}

\author{Orlando Luongo}
\email{orlando.luongo@unicam.it}
\affiliation{Scuola di Scienze e Tecnologie, Universit\`a di Camerino, Via Madonna delle Carceri 9, 62032 Camerino, Italy.}
\affiliation{Dipartimento di Matematica, Universit\`a di Pisa, Largo B. Pontecorvo 5, Pisa, 56127, Italy.}
\affiliation{NNLOT, Al-Farabi Kazakh National University, Al-Farabi av. 71, 050040 Almaty, Kazakhstan.}

\author{Stefano Mancini}
\email{stefano.mancini@unicam.it}
\affiliation{Scuola di Scienze e Tecnologie, Universit\`a di Camerino, Via Madonna delle Carceri 9, 62032 Camerino, Italy.}
\affiliation{Istituto Nazionale di Fisica Nucleare, Sezione di Perugia, Via Alessandro Pascoli 23c, 06123 Perugia, Italy.}

\author{Hernando Quevedo}
\email{quevedo@nucleares.unam.mx}
\affiliation{Instituto de Ciencias Nucleares, Universidad Nacional Autonoma de M\'exico, Mexico D.F. 04510, M\'exico.}
\affiliation{Dipartimento di Fisica and Icra, Universita di Roma “La Sapienza”, Piazzale Aldo Moro 5, I-00185 Roma, Italy.}
\affiliation{Institute for Experimental and Theoretical Physics, Al-Farabi Kazakh National University, Almaty 050040, Kazakhstan.}

\date{\today}

\begin{abstract}
Thermodynamic length is a metric distance between equilibrium thermodynamic states that asymptotically bounds the dissipation induced by a finite time transformation of a thermodynamic system. By means of thermodynamic length, we first evaluate the departures from ideal to real gases in geometric thermodynamics  with and without  Legendre invariance. In particular, we investigate ideal and real gases in the Ruppeiner and geometrothermodynamic formalisms. Afterwards, we formulate a strategy to relate thermodynamic lengths to efficiency of thermodynamic systems in both the aforementioned frameworks in the working assumption of small deviations from ideality. In this respect, we propose a geometric efficiency  definition built up in analogy to quantum thermodynamic systems. We show the result that this efficiency is higher for geometrothermodynamic fluids. Moreover, we stress this efficiency could be used as a  novel  geometric  way  to  distinguish ideal from non-ideal thermal behaviors. In such a way, it could be useful to quantify  deviations from ideality for a variety of real gases. Finally, we discuss the corresponding applications of our recipe to classical thermodynamic systems, noticing that our findings could help geometrically grasping the nature of different metrizations on manifolds of equilibrium thermal states.
\end{abstract}



\maketitle

\section{Introduction}\label{introduzione}

Although thermodynamics represents a well-established theory, various attempts to find a compatible formulation with general relativity (GR) have been so far unsuccessful \cite{termodinamica1,termodinamica1bis,termodinamica1tris}. To focus on this, we can think on how to get the temperature $T$ measured by a speeding observer with $v = const$ if one considers a body at rest with temperature $T_0$. There are several intuitive ways of approaching this issue, for example assuming that the temperature transforms according to special relativity laws, or that the problem is not well-formulated, since if a body moves there is no equilibrium in which one can measure a temperature. In fact, due to the relative movement of observers, a heat flux may appear, leaving the system out of equilibrium. Thus, the temperature $T$ cannot be correctly defined as in classical physics.

In GR the situation is similar: the presence of gravitational field likely implies heat fluxes and so equilibrium is not reached due to gravity. This claim is controversial and, indeed, there are arguments that favor equilibrium even in presence of gravity \cite{termodinamica2,termodinamica2bis,termodinamica2tris}. Thus, a geometric representation of thermodynamics turns out to be essential although not fully understood \cite{termodinamica3,termodinamica3bis,termodinamica3tris}. Geometrical formulations are built up promoting the thermodynamic space of equilibrium states ${\cal E}$ to a differential manifold endowed with a metric structure. This metric is Riemannian and, in this respect, a first proposal consists to endow the space of equilibrium states $\cal{E}$ with metric components obtained from the Hessian matrices of a given thermodynamic potential. In this treatment, if the internal energy $U$ or the negative entropy $-S$ are the potentials, the corresponding metrics are named respectively Weinhold $g^W$ and Ruppeiner $g^R$ metrics \cite{metriche1,metriche1bis,metriche1tris}.

Moreover, Legendre invariance is intimately related to the fact that physical properties of any thermodynamic system do not depend on the selected thermodynamic potential.
In fact, any other thermodynamic potentials can be obtained from the internal energy or entropy via a Legendre transformation. This recipe suggests that a metric based on the Hessian of any thermodynamic potential is a suitable structure for the equilibrium space manifold \cite{metriche2,metriche2bis}. In addition, thermodynamic length is naturally introduced as a consequence of the metric structure above discussed. It is used to compute the distance between equilibrium thermodynamic states \cite{lunghezza1} and it is related to information geometry,  computer simulation \cite{lunghezza3,lunghezza4,lunghezza5.1,lunghezza5.2} and to dissipation for systems near and far from equilibrium \cite{adjunt2}.

The thermodynamic length explicitly depends on the path taken through thermodynamic state space. Consequently, minimum distance paths become geodesics on the Riemannian manifold \cite{lunghezza2}. On the other hand, efficiency of thermodynamic systems is not a well-consolidate notion in relativistic thermodynamics. Above all, a self-consistent theory that matches thermodynamic length with efficiency for given systems is still missing.

In this paper, we consider two accredited relativistic thermodynamic scenarios. In the first scenario, the Legendre invariance does not apply in formulating the thermodynamic metrics, while in the second scenario it does. The first case deals with the standard Fisher-Rao (FR) approach, that is fully recovered by the Ruppeiner's metric. Here, thermodynamic systems can be represented by Riemannian geometry and the statistical properties can be derived accordingly. The second, dubbed geometrothermodynamics (GTD) \cite{adjunt1}, implies Legendre's invariance from the beginning. We demonstrate that in both cases the Fisher-Rao (FR) information metric is recovered for ideal gases. We thus assume real gases, involving the van der Waals equation of state. Here, we compare the thermodynamic lengths between ideal and real gases in the FR scenario first and then in the  GTD formalism. We discuss particular limiting cases in which the van der Waals gas slightly deviates from ideal behavior. Thus, we show  the regions where the thermodynamic length increases and decreases. Afterwards we face the problem of relating the thermodynamic length to thermal efficiency. To do so, we start from the difference between two thermodynamic lengths and from the geodesic shifts. In particular, being in the working assumption of small deviations from ideality, we propose an efficiency measure expressed by means of difference and ratio among different thermal lengths
of ideal and real gases. Differences and discrepancies of our efficiency between ideal and real gases are therefore interpreted. We thus compute the regions of maximum and minimum efficiency in terms of the free parameters of our model for real fluids, fixing one parameter and letting the rest to vary. We finally discuss the physical consequences of our approach in view of possible physical applications.

The paper is structured as follows. In Sect. \ref{sezione2} we analyze the two relativistic thermodynamic scenarios. We first introduce the basic demands of each of them and then we focus on the main differences. In Sect. \ref{sezione3}, we first introduce the idea of thermodynamic length. We then specialize to the FR and GTD cases both ideal and real gases in Sects. \ref{sezione4} and \ref{sezione5}, respectively. We discuss the implications of our results on the different thermodynamic lengths, discussing the main differences among them. In Sect. \ref{sezione6}, we propose our new measure of efficiency. We discuss in detail the consequence of our recipe and we analyze specific cases. We make a comparison between Ruppeiner and GTD efficiencies in Sect. \ref{sezione7}. Finally, conclusions and perspectives are portrayed in Sect. \ref{sezione8}.


\section{Classical thermodynamics and geometry}\label{sezione2}

The empirical laws of classical thermodynamics have been applied in many different scenarios, ranging from microscopic systems up to the Universe itself. Thus, thermodynamics is adopted in several branches of physics, where  one  expects that concepts like non-equilibrium, relativistic, covariant or quantum thermodynamics could be well understood. In this respect, generalizations of the laws of classical thermodynamics are therefore mostly necessary.

For example, the formulation of a \emph{geometric version of classical thermodynamics} seems to be useful in the frameworks of  gravitational and cosmological systems \cite{orl1,orl2,orl3}, in which differential geometry is the main mathematical tool. Indeed, thermodynamics can be incorporated in the description of such systems simply by assuming the validity of equilibrium thermodynamic laws, although there are indications that non-equilibrium effects should also be taken into account. For instance, the dark sector of the Universe can be easily modeled through thermodynamics. In particular, the challenge to understand repulsive gravity in cosmological scenarios seems to a be a prerogative of particular thermodynamic systems \cite{orl4,orl5,orl6}, without passing through the introduction of \emph{ad hoc} dark energy and matter terms.

Differential geometry is a relevant tool as it  describes interactions in terms of geometric concepts. Indeed, the geometric description of gravity and  gauge field theories is based upon  the astonishing principle
{\it field strength = curvature},
first formulated by Einstein in the theory of general relativity. Consequently, in a geometric description of thermodynamics, the curvature is expected to represent the thermodynamic interaction, which is usually interpreted as due to the presence of a potential in the Hamiltonian of the corresponding system.

Essentially, there are two different formalisms that treat thermodynamics in the language of Riemannian differential geometry. The first one is called thermodynamic geometry and consists in introducing Hessian metrics into the space of equilibrium states of the corresponding system. In turn, the Hessian metrics are computed as the second derivatives of a  thermodynamic potential, which can be chosen as the entropy $S$ (Ruppeiner metric), the internal energy $U$ (Weinhold metric) or, in principle, any other thermodynamic potential.
The second formalism, GTD, is characterized by metrics that are Legendre invariant, which takes into account the fact that in classical thermodynamics the properties of a system do not depend on the choice of thermodynamic potential used to describe it.

Our aim is to consider the concepts of thermodynamic length and efficiency in both formalisms and to compare the results in view of possible experimental tests.


\subsection{Thermodynamic geometry}

In classical thermodynamics, a system is usually described by $n$ extensive variables $E^a, \ (a=1,...,n)$, $n$ intensive variables $I_a$, and a thermodynamic potential $\Phi$. The fundamental equation is defined as a function that relates the thermodynamic potential with the extensive variables, $\Phi=\Phi(E^a)$. The equilibrium space ${\cal E}$ is equipped with coordinates $E^a$ so that points of ${\cal E}$ represent  equilibrium states of the system described by $\Phi(E^a)$.  A Hessian metric \cite{adjunt1, metriche1tris}
\be
g^H = \frac{\partial^2\Phi}{\partial E^a \partial E^b} d E^a \otimes d E^b \ ,
\ee
can be introduced in ${\cal E}$, turning it into a Riemannian manifold. The study of this manifold for any thermodynamic system is usually denoted as thermodynamic geometry. In particular, the Weinhold and Ruppeiner metrics are given by
\begin{subequations}\label{Appro}
\begin{align}
g^W & = \frac{\partial^2 U}{\partial E^a\partial E^b}\, dE^a
\otimes dE^b\,,\\
g^R &= - \frac{\partial^2 S}{\partial E^a\partial E^b}\, dE^a \otimes dE^b\,,
\end{align}
\end{subequations}
and are conformally related via
\begin{equation}\label{confo}
g^R =\frac {1}{T} g^W \,.
\end{equation}
where $T$ is the temperature of the system.

The geometric properties of the equilibrium manifold $({\cal E}, g^H)$ are invariant with respect to transformations of the coordinate set $E^a$. The Hessian potential $\Phi$ plays a preference role as the generator of the metric functions and, therefore, cannot be changed arbitrarily.

The Weinhold and Ruppeiner metrics are flat for the ideal gas and curved for other thermodynamic systems with thermodynamic interaction. However, in the case of a system composed of several non-interacting ideal gases, the corresponding equilibrium space turns out to be curved. On the other hand, some black hole configurations have flat equilibrium spaces, although from the point of view of black hole thermodynamics they are characterized by the presence of thermodynamic interaction with phase transitions, (see e.g. \cite{primafcm,FCM,cin1,cin2}).

Ruppeiner metric has interesting and successful applications related to magnetic models \cite{fluido1,fluido2,fluido3} and spin fluids \cite{magnetico1,magnetico2,magnetico3}.

\subsection{GTD}
\label{sec:gtd}

One of the main objective of GTD is to incorporate in the geometric formalism of thermodynamics the fact that the properties of a system do not depend on the choice of thermodynamic potential \cite{ metriche1tris,callen}. This is equivalent to saying that classical thermodynamics is invariant with respect to Legendre transformations \cite{adjunt1}. As mentioned before, in thermodynamic geometry the Hessian potential is not a coordinate of the equilibrium space. To solve this difficulty, we consider the $2n+1$ dimensional phase space ${\cal T}$ with coordinates $Z^A = (\Phi, E^a, I_a)$. Then, a Legendre transformation can be represented as a coordinate transformation of the form: $Z^A\rightarrow Z'^{A}=(\Phi', E'^a, I _a')$,  defined through $\Phi  = \Phi' -  E'^k I'_k, E^i = -  I'_i \,, E^j =  I'_j, 	I_i  = E'^i \,, I_j  = I'_j$, in which $i \in \mathcal I$, $j \in \mathcal J$, $k= 1,...,i$, and $\mathcal I\cup \mathcal J$ is any disjoint decomposition of the set of indices $\{1,...,n\}$.

We point out that when $\mathcal I = \left\{1,\ldots,n\right\}$ and $\mathcal I = \varnothing$, we obtain a total Legendre transformation and the identity,
respectively. Finally, when $\mathcal I\subset \left\{1,\ldots,n\right\}$, we get a partial Legendre transformation.

A metric $G_{AB}$ on ${\cal T}$ is said to be Legendre invariant if its functional form does not change under Legendre transformations. A similar concept of invariance is also used in special relativity, where we demand that the Minkowski metric does not vary under the action of Lorentz transformations. Thus, on ${\cal T}$ we can classify the Legendre invariant metrics as follows:
\begin{subequations}
\begin{align}
G^{^{I/II}} &= (d\Phi - I_a d E^a)^2 + \Lambda(Z^A)(\xi_{ab} E^a I^b) (\chi_{cd} dE^c\otimes dI^d)\,,\label{3a}\\
G^{III} &= (d \Phi - I_a d E^a)^2+\Lambda(Z^A)(E^a I_a)^{2k+1}
	d E^a \otimes d I_a \, ,\label{3b}
\end{align}
\end{subequations}
where $k$ is an integer and $\xi_{ab}=\,\,$diag$(\xi_{11}, ..., \xi_{nn})$ and $\chi_{ab}=$diag$(\chi_{11}, ..., \chi_{nn})$ are constants. Moreover, the conformal factor $\Lambda (Z^A)$ can be choosen as a Legendre invariant function or a constant.
$G^{^{I/II}} $ are invariant under total transformations, whereas $G^{III}$ is also invariant with respect to partial Legendre transformations. Thus, $G^{III}$ is the most general Legendre invariant metric that we have found so far.

Moreover, by virtue of Darboux theorem \cite{28}, there exists a canonical contact one-form $\Theta = d\Phi - I_a d E^a$, satisfying the condition $\Theta \wedge (d\Theta)^n \neq 0$, where $\wedge$  denotes the wedge, or exterior, product on forms commonly used in differential geometry.

In GTD, the equilibrium space ${\cal E}$ is a subspace of ${\cal T}$ that is obtained through the smooth map $\varphi: {\cal E}\to {\cal T}$ such that the pullback of $\varphi$
 annihilates the contact form $\Theta$ and induces a metric on ${\cal E}$, i.e., $\varphi^*(\Theta)=0$ and $g=\varphi^*(G)$. In particular, in the case of the metric $G^{III}$, we obtain that
 \be \label{gGTD}
	g^{III} =  \Lambda \left(E^a \frac{\partial \Phi}{\partial E^a}
	\right)^{2k+1} \frac{\partial^2 \Phi}{\partial E^a \partial E^b}
	\, \d E^a \otimes \d E^b \,.
\ee
A comparison with Weinhold and Ruppeiner metrics with Eq.\eqref{gGTD} shows that the significant departure is due to the conformal term  $ \left(E^a \frac{\partial \Phi}{\partial E^a}	\right)^{2k+1}$.  Notice also that this term has not a definite sign so that the resulting GTD metric could also be pseudo-Riemannian.

In equilibrium thermodynamic fluctuation theory, the components of
Ruppeiner's metric can be interpreted as determining the second fluctuation
moment of the entropy. In GTD, the metrics also have a statistical origin
because they can be expressed in terms of the average and variance of the
microscopic entropy. This has been shown explicitly in \cite{pineda19}.
Moreover, the GTD metrics can also be interpreted as the second fluctuation
moment of a special thermodynamic potential. Indeed, the geometric structure
of GTD allows us to introduce thermodynamic potentials that are not only
related by Legendre transformations, but also by arbitrary transformations,
involving extensive and intensive variables, which are interpreted as
coordinates of the equilibrium space. For each GTD metric, there exists a
particular thermodynamic potential, whose second fluctuation moment is given
in terms of the GTD metric \cite{pineda19}.

Legendre invariance is a property of classical thermodynamics. This
property is incorporated in GTD to guarantee that the geometric and physical
properties of a given system do not depend on the thermodynamic potential
chosen to describe it. In fact, the use of non-Legendre invariant metrics
can lead to inconsistencies and contradictions \cite%
{aman03,aman06A,aman06B,shen07,cai99,sarkar06,medved08,mirza07,quevedo08}.

We notice the freedom in selecting a particular GTD metric, while exhibiting a flavor of arbitrariness, is not particularly relevant in our ranking scheme for two reasons.

First we  acknowledge there is no universal agreement in the literature on the physical
significance of certain geometric quantities in the context of geometric approaches to the physics of gases, see e.g. \cite{lunghezza5.1}. More specifically, both the intensity and sign of the scalar curvature for such gas systems lack a widely shared clear physical interpretation. Moreover, the situation appears to be even more debatable when transitioning from
non-interacting to interacting gases, see e.g. \cite{adj1,adj1ancora}. Second, our ranking scheme is based upon the
concept of \emph{relative differences of geometric quantities}, therefore the
relevance of the absolute numerical value of a geometric quantities becomes
less important.

We here selected the special member of the family
of the GTD metric by setting the conformal factor as $\Lambda=-1$ in Eqs. \eqref{3a}, \eqref{3b} and \eqref{gGTD}, as discussed in Appendix A. This has chosen by imposing computational simplicity and achieving positive definiteness of the metric. Moreover, for the ease of computations, we have chosen a relative integer
parameter, namely $k\in\mathbb{Z}$, by $k=-1$.

In GTD, we interpret the geometric properties of the equilibrium space as describing the thermodynamic properties of the corresponding system, according to the following scheme.
\begin{itemize}
  \item[{\bf 1.}] Curvature of ${\cal E}$ is viewed as thermodynamic interaction.
  \item[{\bf 2.}] Singularities of ${\cal E}$ define phase transitions.
  \item[{\bf 3.}] Certain geodesics of ${\cal E}$ represent quasi-static processes.
\end{itemize}

We point out that in GTD  Legendre invariance can be explicitly shown only at the level of the phase space ${\cal T}$, because a Legendre transformation is equivalent to a coordinate transformation in ${\cal T}$. In the subspace ${\cal E} \subset {\cal T}$, it is not possible to identify Legendre and coordinate transformations. However, we can formally handle Legendre invariance  in ${\cal E}$ by using only quantities that correspond to Legendre invariant quantities in ${\cal T}$. For instance, if we use in ${\cal E}$ the metric $g$, which is derived from the Legendre invariant metric $G$ of ${\cal T}$  derived from $G$
by means of  $g=\varphi^*(G)$, then we say that $g$ is Legendre invariant. This is the case of the thermodynamic length ${\cal L}$ that will be introduced in the next section.

Moreover, a Legendre invariant metric $G$ can be also be subject to diffeomorphisms of ${\cal T}$ that do not affect its geometric properties. This is also true at the level of the equilibrium space ${\cal E}$ for the metric $g=\varphi^*(G)$. This additional diffeomporphism invariance of GTD could have some applications in thermodynamics that are still under investigation. For instance, using the diffeomorphism invariance of ${\cal T}$, we can classify the thermodynamic potential $\Phi$ as follows. If $\Phi$ is taken as the entropy $S$ or the internal energy $U$, we denote $\Phi$ as fundamental potential. A Legendre potential is a potential that is obtained from $\Phi$ by means of a Legendre transformation. Finally, it is also possible to generate diffeomorphic potentials by applying diffeomorphisms on ${\cal T}$. Interestingly, this classification of thermodynamic potentials allows us to find a statistical interpretation for  the GTD metrics \cite{metriche2}.

\section{Thermodynamic length }\label{sezione3}

By using the geometric approach to thermodynamics, it is possible to define the concept of a thermal length of a path $%
\gamma _{\theta }$ in the space of equilibrium states  by \cite{lunghezza4,salam}
\begin{equation}
\mathcal{L}\left( \tau \right) \overset{\text{def}}{=}\int_{0}^{\tau }\sqrt{%
\frac{d\theta ^{a }}{d\xi }g_{ab  }\left( \theta \right) \frac{%
d\theta ^{b }}{d\xi }}d\xi \text{,}  \label{length}
\end{equation}%
where the parameter $\theta $ depends on an affine parameter $\xi $ with $%
0\leq \xi \leq \tau $, and $g_{ab }(\theta )$ is the thermodynamic
metric tensor. The thermal length, $\mathcal{L}(\tau )$,  refers to the path $%
\gamma _{\theta }$ and represents a measure of the cumulative root-mean-square
deviations along the path $\gamma _{\theta }$ \cite{lunghezza3}. In the case of thermodynamic geometry, it is possible to formulate $\mathcal L(\tau)$ in terms of both energy and entropy, leading to slight differences; see e.g. \cite{salamon85}.

The use of thermodynamic length is wide and it can be applied in several observable contexts. For example, it is possible to relate it either to the existence probability of a given system or to quantifying minimum entropy production paths emerging from quantum mechanical evolution of relevance in continuous-time quantum searching \cite{cafaro2020pre}. The way to parametrize Eq. \eqref{length} permits
to define a point-to-point distance being the shortest curve distance, i.e. geodesics, analogous to  straight lines in a curved space. This is relevant to point out  because of the connection to fluctuations. The curve lengths are measured by the number of natural fluctuations along the path. Our aim in this work  is to compute Eq. \eqref{length}, using different thermodynamic metrics.

Thus, we here consider two main physical scenarios in which we apply the definition of thermal length, namely, the ideal gas and the van der Waals gas,  the most famous extension to the case of real gases. Our findings  should emphasize  the physical departures that the van der Waals gas shows with respect to the ideal gas. In particular, departing from ideality, the van der Waals gas model contains two non-ideal ingredients: $1)$ a weak long-range attraction among molecules and $2)$ a strong
short-range repulsion among molecules of the gas.

More specifically, the
substance-specific constants $a$ and $b$ are measures of the long-range
attraction and the short-range repulsion, respectively. Comparing Eqs.
\ref{entropyideal} and \ref{entropyreal} of Appendix B, we note that the presence of a
non-vanishing $b$ tends to decrease the entropy of the gas while the presence
of a nonzero $a$ tends to increase it. A similar behavior is expected in the framework of the thermal length, by varying the parameters $a$ and $b$. In fact, when $b\neq0$, the
entropy decreases  since each mole of gas occupies an effective volume
of $b/N_{A}$ and so the van der Waals particles cannot occupy the same space. Thus,
the decrease of the entropy when $b\neq0$ can be understood in terms of a
smaller number of position states in a van der Waals gas than in an ideal gas.
When $a\neq0$, the entropy increases because more velocity states are
available for the same internal energy due to attractive interactions among
particles. Indeed, for an ideal and a van der Waals gas we have $0\leq
v_{\mathrm{ideal}}\lesssim\left[  \left(  2K\right)  /m\right]  ^{1/2}$ and
$0\leq v_{\mathrm{vdW}}\lesssim\left[  \left(  2K+N\frac{a}{b}\right)
/m\right]  ^{1/2}$, respectively, with $K$ denoting pure kinetic energy. Thus,
the increase of the entropy when $a\neq0$ can be explained in terms of a
larger number of velocity states in a van der Waals gas than in an ideal gas.

By virtue of the above considerations, we expect that in the two formalisms that we investigate the thermal lengths for the van der Waals could be  larger or smaller than the one computed for ideal gases for quite negligible $a$ and  $b$. We investigate in which cases this is true and we show that the van der Waals length is larger than the ideal gas length as $a\ll1$ and $b\ll1$ in thermodynamic geometry and in GTD.

\section{Ruppeiner  geometry}
\label{sezione4}

In Appendix B, we derive the fundamental equation for the ideal gas, using the entropy as thermodynamic potential. It is then straightforward to calculate the corresponding Ruppeiner metric,
\be
g^R_{\mathrm{ideal}}
=\frac{nC_{V}%
}{U^{2}}dU^{2}+\frac{nR}{V^{2}}dV^{2}\text{,}
\ee
whose curvature tensor vanishes identically. This shows that the corresponding equilibrium space is flat due to the lack of thermodynamic interaction.

We now work out  the thermodynamic length  by using the Ruppeiner metric. We thus assume that  $\theta^{a}\left(  \xi\right)  $ is a path satisfying the geodesic equations. In the case of the ideal gas, evaluating the thermodynamic length along the geodesic paths $\theta^{a
}=\theta^{a}\left(  \xi\right)  $, and considering the results of Appendix C, we have%
\begin{equation}
\mathcal{L}_{\mathrm{ideal}}^{\text{R}}\left(  \tau\right)  =\left[  nC_{V}\left(
\frac{\dot{U}_{0}}{U_{0}}\right)  ^{2}+nR\left(  \frac{\dot{V}_{0}}{V_{0}%
}\right)  ^{2}\right]  ^{1/2}\tau\text{,} \label{lideal1}%
\end{equation}
where $U_{0}\overset{\text{def}}%
{=}U\left(  0\right)  $, $V_{0}\overset{\text{def}}{=}V\left(  0\right)  $,
$\dot{U}_{0}\overset{\text{def}}{=}\left(  dU/d\xi\right)  _{\xi=0}$, and
$\dot{V}_{0}\overset{\text{def}}{=}\left(  dV/d\xi\right)  _{\xi=0}$.

The geodesic equations are formally $
\frac{d^{2}\theta^{a}}{d\xi^{2}}+\Gamma_{bc}^{a}\frac{d\theta^{b}%
}{d\xi}\frac{d\theta^{c}}{d\xi}=0$ and give
\begin{equation}
\frac{d^{2}U}{d\xi^{2}}-\frac{1}{U}\left(  \frac{dU}{d\xi}\right)
^{2}=0,\quad \frac{d^{2}V}{d\xi^{2}}-\frac{1}{V}\left(  \frac{dV}{d\xi
}\right)  ^{2}=0\text{.} \label{geq}%
\end{equation}

Then, integrating Eqs. \eqref{geq}  yields:
$U\left(  \xi\right)  =U_{0}e^{\frac{\dot{U}_{0}}{U_{0}}\xi}$ and $V\left(  \xi\right)  =V_{0}e^{\frac{\dot{V}_{0}}{V_{0}}\xi}$. If we perform a suitable change of parameters, geodesic paths become
straight lines, e.g. consider,  $
\xi\rightarrow\eta=\eta\left(  \xi\right)  \overset{\text{def}}{=}%
e^{\frac{\dot{U}_{0}}{U_{0}}\xi}\text{, that is, }\xi=\xi\left(  \eta\right)
\overset{\text{def}}{=}\frac{U_{0}}{\dot{U}_{0}}\log\left(  \eta\right)$, we have the geodesic equation for $\tilde{U}\left(  \eta\right)
\overset{\text{def}}{=}U\left(  \xi\left(  \eta\right)  \right)  $ and we recover Eq. \eqref{lideal1}.

In the case of the van der Waals gas, we can evaluate the thermodynamic length along the geodesic paths $\theta^{a
}=\theta^{a}\left(  \xi\right)  $ to get
\begin{widetext}
\begin{equation}\label{10}
\mathcal{L}_{\mathrm{vdW}}^{\text{R}}\left(  \tau\right)  =\left\{  \frac{nC_{V}V_{0}%
^{2}\dot{U}_{0}^{2}}{\left(  an^{2}+U_{0}V_{0}\right)  ^{2}}-\frac
{2an^{3}C_{V}\dot{U}_{0}\dot{V}_{0}}{\left(  an^{2}+U_{0}V_{0}\right)  ^{2}%
}+\left[  \frac{nR}{\left(  V_{0}-nb\right)  ^{2}}-\frac{an^{3}C_{V}}%
{V_{0}^{2}}\frac{\left(  an^{2}+2U_{0}V_{0}\right)  }{\left(  an^{2}%
+U_{0}V_{0}\right)  ^{2}}\right]  \dot{V}_{0}^{2}\right\}  ^{1/2}\tau\text{.}%
\end{equation}
\end{widetext}

In Ruppeiner geometry, to compute the explicit components of the
metric, one can use only extensive variables. The reason is that Ruppeiner
metric is the Hessian of the entropy with respect to the extensive variables
only. In GTD, the situation is different. Due to Legendre invariance imposed \emph{a priori} over the metric, one
can use any thermodynamic potential and any combination of extensive and
intensive variables that are related by means of Legendre transformations.
The structure of GTD guarantees that the geometric properties of the
equilibrium space and the physical properties of the corresponding system do
not depend on the choice of potential and variables.

\subsection{Comparison between ideal and real gases}

To compare ideal and real gases in thermodynamic geometry, we assume reasonable values for the free terms entering the above outcomes. Thus, we assume $n=1$, $U_{0}=1$, $V_{0}
=1$, $\dot{U}_{0}=1$,
and $\dot{V}_{0}=1$ and we get
\begin{subequations}\label{nove}
\begin{align}
\mathcal{L}_{\mathrm{ideal}}^{\text{R}}\left(  \tau\right)&=\left[  C_{V}+R\right]
^{\large{\frac{1}{2}}}\tau\,, \label{lideal}\\
\mathcal{L}_{\mathrm{vdW}}^{\text{R}}\left(  \tau\right)  &=\left[  \frac{5-\left(
a+2\right)  ^{2}}{\left(  1+a\right)  ^{2}}C_{V}+\frac{1}{\left(  1-b\right)
^{2}}R\right]  ^{\frac{1}{2}}\tau\,. \label{lreal}%
\end{align}
\end{subequations}
We observe that for any $a\in\left(  0\text{, }1\right)  $ and $b\in\left(
0\text{, }1\right)  $, we have $
\frac{5-\left(  a+2\right)  ^{2}}{\left(  1+a\right)  ^{2}}\leq1\text{, and
}\frac{1}{\left(  1-b\right)  ^{2}}\geq1$,
respectively.

In terms of Eq. \eqref{lreal}, we note that the presence of non-vanishing van der Waals parameters $a$ and $b$ exhibit competing effects. In particular, weak long-range attraction between molecules tends to shrink the
thermodynamic length while strong short-range repulsion seems to favor the growth of the thermodynamic length. Thus, at some intermediate pressure values, the two corrections have opposite influences and we have a more definite
situation at low and high pressures, respectively. Specifically, at low pressures, the correction for intermolecular attraction related to the parameter $a$ is more important than the one for the molecular volume, $b$. In such a scenario, we
have from Eqs. (\ref{lideal}) and (\ref{lreal}) that%
\begin{equation}\label{realapprossimatoFR1}
\mathcal{L}_{\mathrm{vdW}}^{\text{R}}\left(  \tau\right)  \overset{b\rightarrow0}%
{\approx}\left[  \frac{5-\left(  a+2\right)  ^{2}}{\left(  1+a\right)  ^{2}%
}C_{V}+R\right]  ^{\frac{1}{2}}\tau\,.
\end{equation}

At high pressures and small volumes, instead, the correction for the volume of
molecules becomes more important because molecules themselves are
incompressible and constitute a
non-negligible fraction of the total volume of
the gas. Here, we have
\begin{equation}\label{realapprossimatoFR2}
\mathcal{L}_{\mathrm{
vdW}}^{\text{R}}\left(  \tau\right)  \overset{a\rightarrow0}%
{\approx}\left[  C_{V}+\frac{1}{\left(  1-b\right)  ^{2}}R\right]  ^{\frac{1}{2}}%
\tau.
\end{equation}
Thus, in the approximations $a\rightarrow0$ and $b\rightarrow0$, respectively, we have
\begin{subequations}\label{set21}
\begin{align}
&\mathcal{L}_{\mathrm{vdW}}^{\text{R}}\left(  \tau\right)\geq\left[  C_{V}+R\right]^{1/2}\tau=\mathcal{L}_{\mathrm{ideal}}^{\text{R}}\left(  \tau\right)  \,,\\
&\mathcal{L}_{\mathrm{vdW}}^{\text{R}}\left(  \tau\right)\leq\left[  C_{V}+R\right]^{1/2}\tau=\mathcal{L}_{\mathrm{ideal}}^{\text{R}}\left(  \tau\right).
\end{align}
\end{subequations}

Hence, in the case of Ruppeiner geometry, considering the two distinct approximations, i.e. $a\rightarrow0$ and $b\rightarrow0$, we infer opposite behaviors of the thermodynamic lengths, respectively. In fact, we can recast Eqs. \eqref{set21} by

\begin{subequations}\label{rese1}
\begin{align}
\mathcal{L}_{\mathrm{vdW}}^{\text{R}}\left(  \tau\right)\overset{b\ll1}{\approx}\left(
\allowbreak\allowbreak1+\frac{2}{5}b\right)  \mathcal{L}_{\mathrm{ideal}%
}^{\text{R}}\left(  \tau\right)  \geq\mathcal{L}_{\mathrm{ideal}%
}^{\text{R}}\left(  \tau\right), \\
\mathcal{L}_{\mathrm{vdW}}^{\text{R}}\left(  \tau\right)  \overset{a\ll
1}{\approx}\left(  \allowbreak1-\frac{9}{5}a\right)  \mathcal{L}%
_{\mathrm{ideal}}^{\text{R}}\left(  \tau\right)  \leq\mathcal{L}%
_{\mathrm{ideal}}^{\text{R}}\left(  \tau\right)
\end{align}
\end{subequations}
respectively with $a=0$ and $b=0$. The behaviors of Eqs. (\ref{nove}-\ref{realapprossimatoFR1}) are portrayed in Fig. 1.

\begin{figure}\label{figura1}
\centering
\includegraphics[width=0.99\columnwidth,clip]{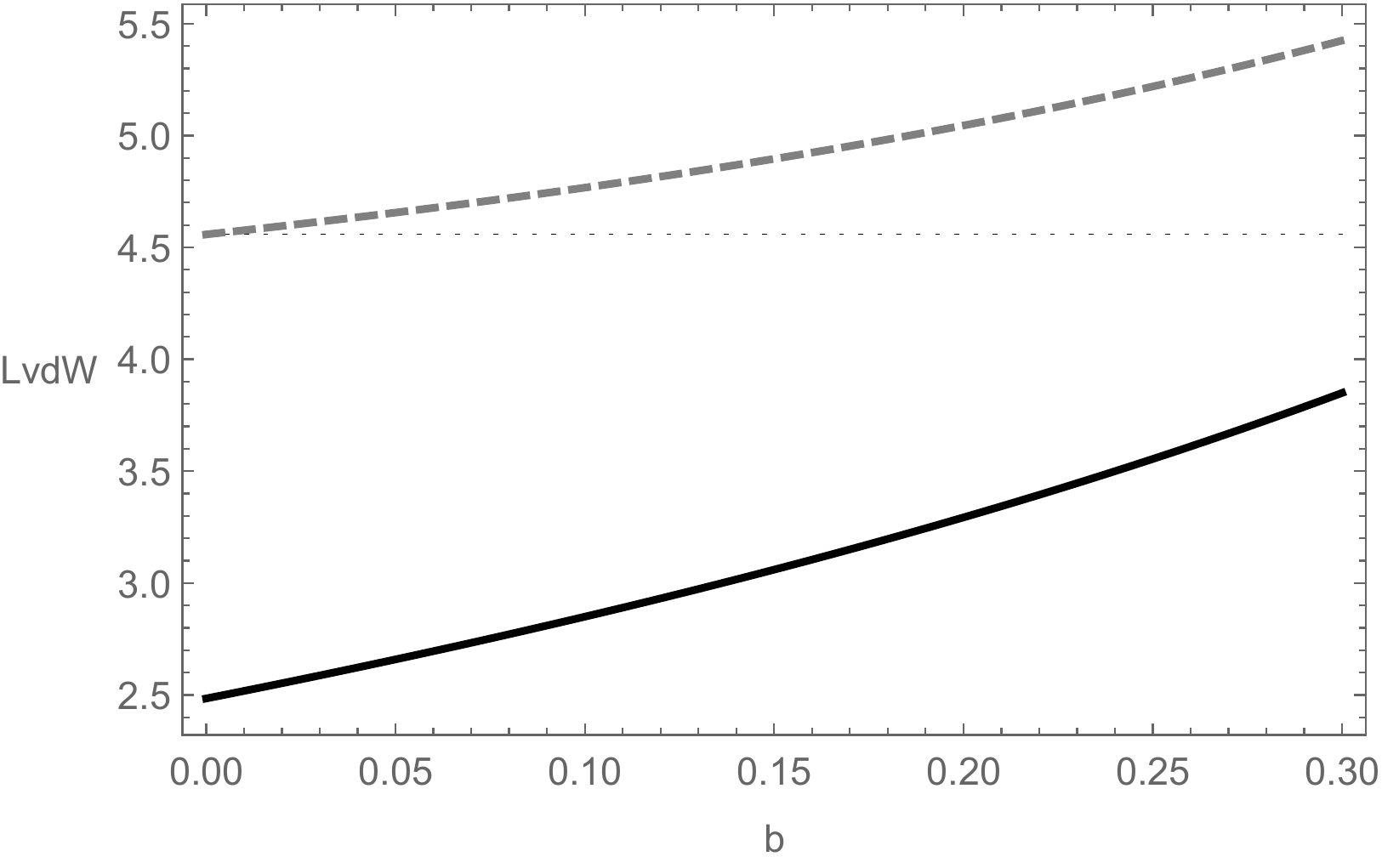}
\includegraphics[width=0.99\columnwidth,clip]{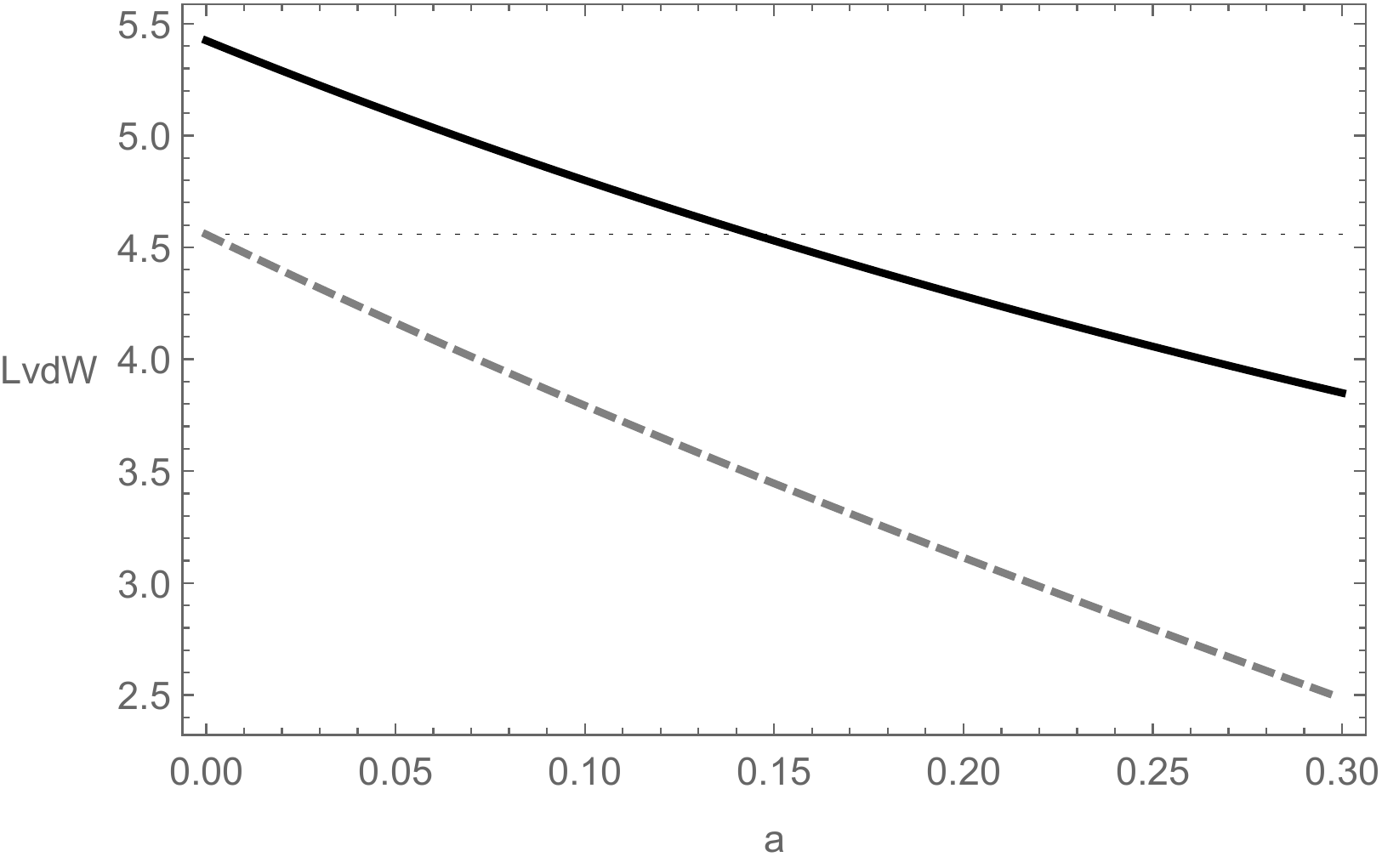}
\caption{\textit{Top panel}: plot of the  thermodynamic length $\mathcal L^{\text{R}}_{\mathrm{vdW}}$ of Eq. \eqref{lreal} as a  function of  $b$. The plot refers to two indicative choices: $a=0$ and $a=0.3$, corresponding to the gray and black  lines, respectively. The ideal gas, reported in Eq. \eqref{lideal}, is the dotted constant line. \textit{Bottom panel}: the same quantity as above, this time as function of $a$ with $b$ fixed to $0$ and $0.3$. For both plots we considered the indicative bounds $C_V=1.5R$, compatible with monoatomic gas specific heat value, and $\tau=1$.}
\end{figure}

\subsection{Ruppeiner thermodynamic lengths}

Without loss of generality, our
particular choice of initial conditions is merely dictated for illustrative
purposes. However, more general physical boundary conditions could have been
taken into consideration. Here, we show that with arbitrary choice of such conditions the functional behaviors of thermodynamic lengths do not change. In particular, for arbitrary physical initial conditions, we have for $a=0, b\ll 1$
\begin{align}
\mathcal{L}_{\mathrm{vdW}}^{\text{R}}\left(\tau\right)
&\approx \left[ \frac{3}{2}k\left( \frac{\dot{U}_{0}}{U_{0}}\right)
^{2}+k\left( \frac{\dot{V}_{0}}{V_{0}}\right) ^{2}\left( 1+2\frac{b}{V_{0}}%
\right) \right] ^{1/2}\tau\nonumber\\
&\geq \mathcal{L}_{\mathrm{ideal}}^{\text{R}}\left( \tau \right) \text{,}
\end{align}%
whereas for $b=0, a\ll 1$, we get

\begin{align}\label{EQUA}
\mathcal{L}_{\mathrm{vdW}}^{\text{R}}\left(\tau\right) &\approx \Bigg\{
\frac{3}{2}k\left( \frac{\dot{U}_{0}}{U_{0}}\right) ^{2}\allowbreak \left(
1-2\frac{a}{U_{0}V_{0}}\right) +k\left( \frac{\dot{V}_{0}}{V_{0}}\right)
^{2}+ \nonumber\\
&-\left( \frac{3ak}{U_{0}^{2}V_{0}^{2}}\right) \left[ \dot{U}_{0}\dot{V}%
_{0}+U_{0}V_{0}\left( \frac{\dot{V}_{0}}{V_{0}}\right) ^{2}\right]\Bigg\}^{1/2}\tau\nonumber\\
&\leq \mathcal{L}_{\mathrm{ideal}}^{\text{R}}\left(\tau\right)\text{.}
\end{align}

We point out the generality of the inequality in Eq. \eqref{EQUA}, which is clearly valid when
all the boundary conditions are strictly positive.
This type of conditions can be used, for example, to describe a thermal
process where the gas expands during its evolution with a starting positive
rate of change of its internal energy. For the sake of illustrative purposes,
we will  work within these  conditions in what follows.

Further, we emphasize that to quantify the maximum efficiency in a heat engine subject to losses, the minimization of entropy production during a complete cycle is performed considering the minimization
procedure in each consecutive branch defining the cycle subject to specific
boundary conditions. As pointed out in Ref. \cite{adj2}, the optimization procedure depends on the equations of state of the working fluid and becomes analytically
intractable even for an ideal gas in a cylinder if one attempts to perform an
optimization subject to general constraints. Moreover, when transitioning
between two consecutive branches, different initial conditions can be assumed.
In our proposed geometric thermodynamics analysis, our choice of initial
conditions corresponds to choosing specific initial conditions for each
branch of a thermodynamic cycle.

\section{GTD geometry}
\label{sezione5}

In the case of ideal or real gases, the most general GTD metric yields

\begin{align}
 g^{\text{GTD}}&=-\left(  U\frac{\partial S}{\partial
U}\right)  ^{2k+1}\frac{\partial^{2}S}{\partial U^{2}}dU^{2}\\
&-\left[  \left(
U\frac{\partial S}{\partial U}\right)  ^{2k+1}+\left(  V\frac{\partial
S}{\partial V}\right)  ^{2k+1}\right]  \frac{\partial^{2}S}{\partial U\partial
V}dUdV\nonumber\\
&-\left(  V\frac{\partial S}{\partial V}\right)  ^{2k+1}\frac
{\partial^{2}S}{\partial V^{2}}dV^{2}.\nonumber \label{30}%
\end{align}
Thus, considering Eqs. \ref{entropyideal} and \ref{entropyreal} of Appendix B, we consider below the cases for ideal and real gases, respectively.

Moreover, physically meaningful geodesic paths connecting two equilibrium thermodynamic states on the equilibrium manifold equipped with the GTD metric represent quasi-static processes, i.e. processes characterized by \emph{infinite slowness}, satisfying the laws of thermodynamics \cite{metriche2bis}. A quasi-static process does not need to be reversible and, indeed, quasi-static processes involving entropy production are irreversible. When the evolution along the geodesic path occurs with no entropy increase, the thermodynamic geodesic is called an adiabatic geodesic and it describes a reversible quasi-static process. When the evolution along the geodesic occurs with entropy increase, one has a non-adiabatic geodesic that describes an irreversible quasi-static process.

In any case, either reversible or irreversible, geodesics have a definite direction associated with the direction in which the entropy increases and correspond to minimally dissipative processes where the optimum cooling (geodesic) paths happen with maximum reversibility, i.e., at minimum entropy production \cite{adj23}.

\subsection{Comparison between ideal and real gases}

In the case of ideal gas, by virtue of the results of Appendices C and D, we immediately get
\begin{equation}
g_{\mathrm{ideal}}^{\text{GTD}}=\frac{(nC_{V})^{2k+2}%
}{U^{2}}dU^{2}+\frac{\left(  nR\right)  ^{2k+2}}{V^{2}}dV^{2}\text{,}%
\end{equation}
a metric that leads to a vanishing curvature, independently of the value of the constant $k$.

The corresponding thermodynamic length can be evaluated by following the line of reasoning outlined earlier and we get%
\begin{equation}\label{20}
\mathcal{L}_{\mathrm{ideal}}^{\text{GTD}}\left(  \tau\right)  =\left[  \left(
nC_{V}\right)  ^{2k+2}\left(  \frac{\dot{U}_{0}}{U_{0}}\right)  ^{2}+\left(
nR\right)  ^{2k+2}\left(  \frac{\dot{V}_{0}}{V_{0}}\right)  ^{2}\right]
^{1/2}\tau\text{.}%
\end{equation}
If, in addition, we set $k=-1$ for simplicity, we
obtain%
\begin{equation}
\mathcal{L}_{\mathrm{ideal}}^{\text{GTD}}\left(  \tau\right)  =\left[  \left(
\frac{\dot{U}_{0}}{U_{0}}\right)  ^{2}+\left(  \frac{\dot{V}_{0}}{V_{0}%
}\right)  ^{2}\right]  ^{1/2}\tau\text{.}%
\end{equation}
Assuming, as above $U_{0}\overset{\text{def}}{=}U\left(  0\right)  =1$, $V_{0}%
\overset{\text{def}}{=}V\left(  0\right)  =1$, $\dot{U}_{0}\overset
{\text{def}}{=}\left(  dU/d\xi\right)  _{\xi=0}=1$, and $\dot{V}_{0}%
\overset{\text{def}}{=}\left(  dV/d\xi\right)  _{\xi=0}=1$,  the above thermodynamic length becomes%
\begin{equation}
\mathcal{L}_{\mathrm{ideal}}^{\text{GTD}}\left(  \tau\right)  =2^{1/2}%
\tau\text{.} \label{compare1}%
\end{equation}

An analogous procedure can be applied to the van der Waals gas. From Eq. \ref{entropyreal} of Appendix B, which corresponds to the fundamental equation of the van der Waals gas, we obtain%

\begin{subequations}\label{31}%
\begin{align}
\left(  U\frac{\partial S}{\partial U}\right)  ^{2k+1}  &  =\left(
nC_{V}\frac{U}{U+n^{2}\frac{a}{V}}\right)  ^{2k+1},\\
\left(  V\frac{\partial S}{\partial V}\right)  ^{2k+1}  &  =\left(  nR\frac
{V}{V-nb}-nC_{V}\frac{n^{2}\frac{a}{V}}{U+n^{2}\frac{a}{V}}\right)^{2k+1},
\end{align}
\end{subequations}

\noindent and since %
\begin{subequations}
\begin{align}
\frac{\partial^{2}S}{\partial U^{2}}  &  =-nC_{V}\frac{V^{2}}{\left(
an^{2}+UV\right)  ^{2}}\text{,}\\
\frac{\partial^{2}S}{\partial U\partial V}  &  =\frac{an^{3}C_{V}}{\left(
an^{2}+UV\right)  ^{2}}\text{,}\\
\frac{\partial^{2}S}{\partial V^{2}}  &  =-\frac{nR}{V^{2}}\frac{V^{2}%
}{\left(  V-bn\right)  ^{2}}+\frac{an^{3}C_{V}}{V^{2}}\frac{\left(
an^{2}+2UV\right)  }{\left(  an^{2}+UV\right)  ^{2}}\text{.}
\end{align}\label{32}%
\end{subequations}

by virtue of  Eqs. (\ref{31}) and (\ref{32}), the  GTD metric reads
\begin{widetext}
\begin{align}
g_{\mathrm{vdW}} ^{\text{GTD}}  &  =\left(  nC_{V}%
\frac{U}{U+n^{2}\frac{a}{V}}\right)  ^{2k+1}\frac{nC_{V}V^{2}}{\left(
an^{2}+UV\right)  ^{2}}dU^{2}\nonumber\\
& \nonumber\\
&  -\left[  \left(  nC_{V}\frac{U}{U+n^{2}\frac{a}{V}}\right)  ^{2k+1}+\left(
nR\frac{V}{V-nb}-nC_{V}\frac{n^{2}\frac{a}{V}}{U+n^{2}\frac{a}{V}}\right)
^{2k+1}\right]  \frac{an^{3}C_{V}}{\left(  an^{2}+UV\right)  ^{2}%
}dUdV\nonumber\\
& \nonumber\\
&  -\left(  nR\frac{V}{V-nb}-nC_{V}\frac{n^{2}\frac{a}{V}}{U+n^{2}\frac{a}{V}%
}\right)  ^{2k+1}\left[  -\frac{nR}{V^{2}}\frac{V^{2}}{\left(  V-bn\right)
^{2}}+\frac{an^{3}C_{V}}{V^{2}}\frac{\left(  an^{2}+2UV\right)  }{\left(
an^{2}+UV\right)  ^{2}}\right]  dV^{2}\text{.}%
\end{align}
\end{widetext}

To simplify our computation, we set $n=1$, $k=-1$, $C_{V}=(3/2)R$
(monoatomic gas). Then, we have
\begin{align}
g_{\mathrm{vdW}} ^{\text{GTD}}  &  =\frac{1}%
{U^{2}+a\frac{U}{V}}dU^{2}\nonumber\\
&+\frac{a}{aUV+U^{2}V^{2}}\frac{3ab-aV-3bUV+5UV^{2}%
}{3ab-aV+2UV^{2}}dUdV\nonumber\\
& \nonumber\\
&  -\frac{a\left(  a+2UV\right)  \left(  3b^{2}-6bV+V^{2}\right)  -2U^{2}%
V^{4}}{V^{2}\left(  a+UV\right)  \left(  V-b\right)  \left(  3ab-aV+2UV^{2}%
\right)  }dV^{2}\text{.}%
\end{align}

The thermodynamic length can now be computed with the above metric. Thus, following the line of reasoning outlined above, we get%
\begin{align}\label{conlu}
&\left(\frac{\mathcal{L}_{\mathrm{vdW}}^{\text{GTD}}\left(  \tau\right)}{\tau}\right)^2=\frac{\dot{U}_{0}^{2}}{U_{0}^{2}+a\frac{U_{0}}{V_{0}}}\nonumber\\\,\nonumber\\%
&-\frac{a}{aU_{0}%
V_{0}+U_{0}^{2}V_{0}^{2}}\frac{3ab-aV_{0}-3bU_{0}V_{0}+5U_{0}V_{0}^{2}%
}{3ab-aV_{0}+2U_{0}V_{0}^{2}}\dot{U}_{0}\dot{V}_{0}\nonumber\\
\,\nonumber\\
&-\frac{a\left(  a+2U_{0}V_{0}\right)  \left(  3b^{2}-6bV_{0}+V_{0}^{2}\right)
-2U_{0}^{2}V_{0}^{4}}{V_{0}^{2}\left(  a+U_{0}V_{0}\right)  \left(
V_{0}-b\right)  \left(  3ab-aV_{0}+2U_{0}V_{0}^{2}\right)  }\dot{V}_{0}^{2}%
\end{align}
Assuming the same value of the free terms entering Eq. \eqref{conlu},  the quantity $\mathcal{L}%
_{\mathrm{vdW}}^{\text{GTD}}\left(  \tau\right)  $ becomes
\begin{equation}
\mathcal{L}_{\mathrm{vdW}}^{\text{GTD}}\left(  \tau\right)  =2^{1/2}\left[
\frac{4a+b+6ab^{2}-a^{2}b-12ab-2}{\left(  1-b\right)  \left(  1+a\right)
\left(  a-3ab-2\right)  }\right]  ^{1/2}\tau\text{.} \label{compare2}%
\end{equation}
Note that the above expression reduces to $\mathcal L_{ideal}^{\text{GTD}}$ as $a\rightarrow0$ and $b\rightarrow0$.

Comparing Eqs. (\ref{compare1}) and (\ref{compare2}), we finally get%
\begin{equation}
\frac{\mathcal{L}_{\mathrm{vdW}}^{\text{GTD}}\left(  \tau\right)
}{\mathcal{L}_{\mathrm{ideal}}^{\text{GTD}}\left(  \tau\right)  }=\left[
\frac{4a+b+6ab^{2}-a^{2}b-12ab-2}{\left(  1-b\right)  \left(  1+a\right)
\left(  a-3ab-2\right)  }\right]  ^{1/2}\text{.}%
\end{equation}

It is interesting to check what changes occur in two intriguing limiting cases that correspond to a first subcase, $b\rightarrow0$, where we infer
\begin{equation}
\mathcal{L}_{\mathrm{vdW}}^{\text{GTD}}\left(  \tau\right)  =\left[
\frac{2-4a}{\left(  1+a\right)  \left(  2-a\right)  }\right]  ^{1/2}%
\mathcal{L}_{\mathrm{ideal}}^{\text{GTD}}\left(  \tau\right)  \text{,}%
\end{equation}
and in the subcase $a\rightarrow0$, where we find
\begin{equation}\label{abraca}
\mathcal{L}_{\mathrm{vdW}}^{\text{GTD}}\left(  \tau\right)  =\left[
\frac{2-b}{2\left(  1-b\right)  }\right]  ^{1/2}\mathcal{L}_{\mathrm{ideal}%
}^{\text{GTD}}\left(  \tau\right) ,%
\end{equation}
in agreement with the analysis based on the  Ruppeiner metric.

The behaviors of the thermal lengths in the GTD case are portrayed in Fig. 2, where the above limits, i.e. Eqs. \eqref{abraca}, are recovered. Moreover, for small values of the interaction parameters, respectively with $a=0$ and $b=0$, the following relationships hold:

\begin{figure}\label{figura2}
\centering
\includegraphics[width=0.99\columnwidth,clip]{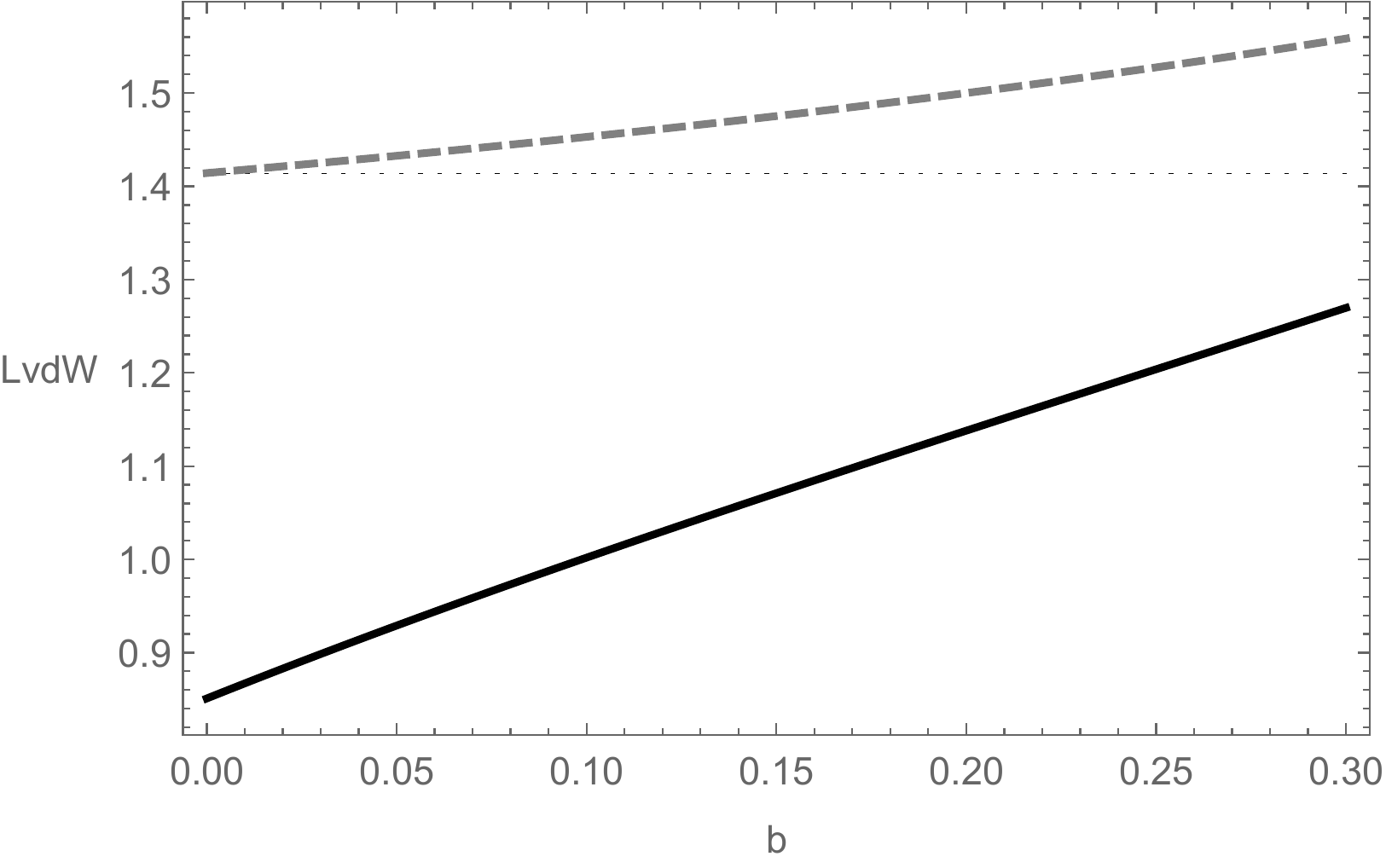}
\includegraphics[width=0.99\columnwidth,clip]{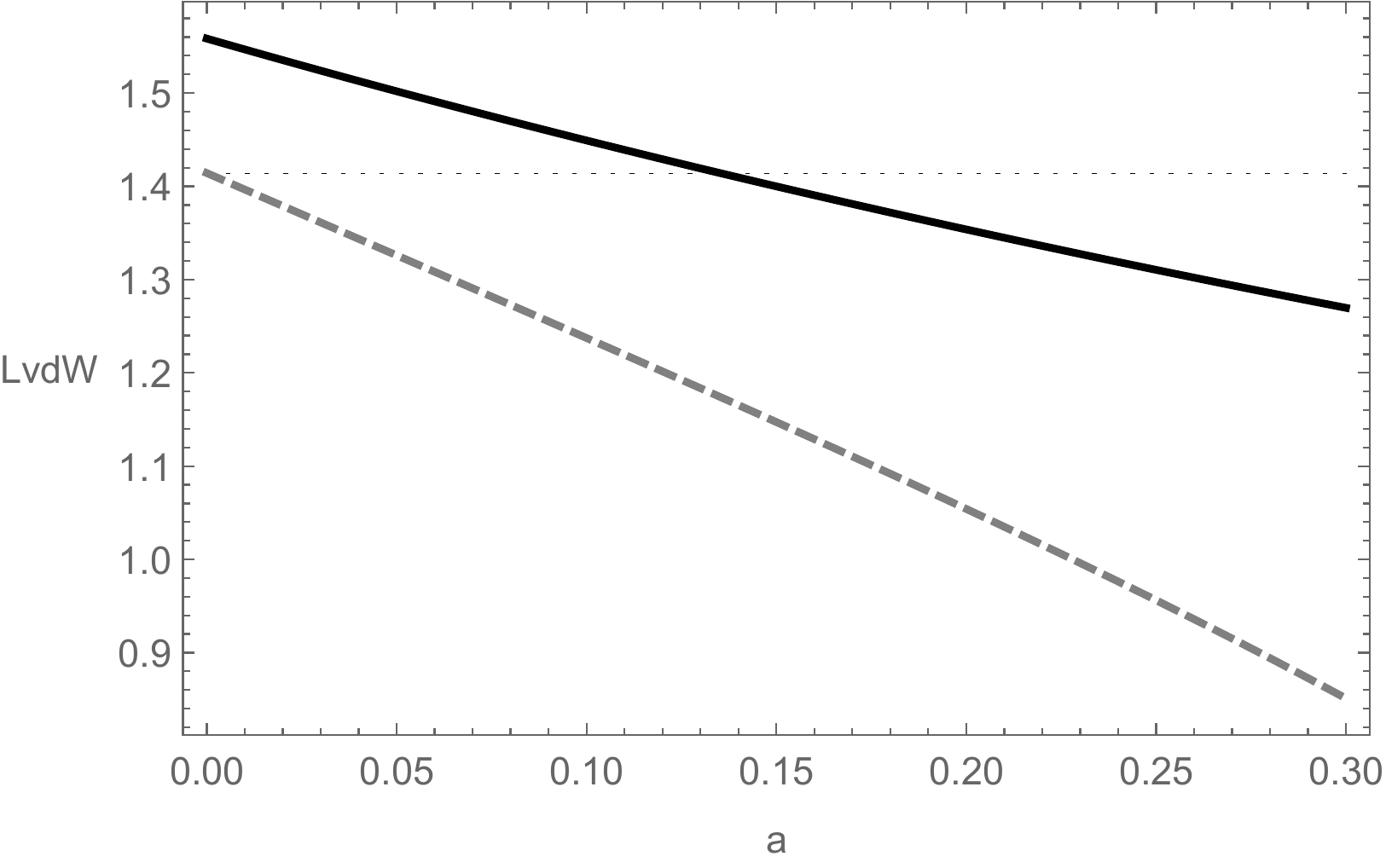}
\caption{\textit{Top panel}: plot of the  thermodynamic length $\mathcal L^{\text{GTD}}_{\mathrm{vdW}}$ of Eq. \eqref{compare2} as a function of  $b$. The plot refers to two indicative choices: $a=0$ and $a=0.3$, respectively the gray and black  lines. The ideal gas, reported is the straight dotted line. \textit{Bottom panel}: the same quantity as above, this time as function of $a$ with $b$ fixed to $0$ and $0.3$. In both the graphics, we conventionally assumed $\tau=1$.}
\end{figure}

\begin{subequations}\label{rese2}
\begin{align}
\mathcal{L}_{\mathrm{vdW}}^{\text{GTD}}\left(  \tau\right)\overset{b\ll1}{\approx}\left(
\allowbreak\allowbreak1+\frac{1}{4}b\right)  \mathcal{L}_{\mathrm{ideal}%
}^{\text{GTD}}\left(  \tau\right)  \geq\mathcal{L}_{\mathrm{ideal}%
}^{\text{GTD}}\left(  \tau\right), \\
\mathcal{L}_{\mathrm{vdW}}^{\text{GTD}}\left(  \tau\right)  \overset{a\ll
1}{\approx}\left(  \allowbreak1-\frac{5}{4}a\right)  \mathcal{L}%
_{\mathrm{ideal}}^{\text{GTD}}\left(  \tau\right)  \leq\mathcal{L}%
_{\mathrm{ideal}}^{\text{GTD}}\left(  \tau\right).
\end{align}
\end{subequations}

\subsection{GTD thermodynamic lengths}

In analogy to what we found for the Ruppeiner case above, we stress our particular choice of initial
conditions is just illustrative. In the GTD case, we observe that for arbitrary physical conditions, we have for $a=0, b\ll1$%
\begin{align}
\mathcal{L}_{\mathrm{vdW}}^{\text{GTD}}\left( \tau \right)&\approx \left[ \left( \frac{\dot{U}_{0}}{U_{0}}\right)
^{2}+\left( \frac{\dot{V}_{0}}{V_{0}}\right) ^{2}\left( 1+\frac{b}{V_{0}}%
\right) \right] ^{1/2}\tau \nonumber\\
&\geq\mathcal{L}_{\mathrm{ideal}}^{\text{GTD}%
}\left( \tau \right) \text{,}
\end{align}%
and for $b=0, a\ll1$
\begin{align}
\mathcal{L}_{\mathrm{vdW}}^{\text{GTD}}\left( \tau \right)&\approx \left[ \left( \frac{\dot{U}_{0}}{U_{0}}\right)
^{2}\left( 1-\frac{a}{U_{0}V_{0}}\right) +\left( \frac{\dot{V}_{0}}{V_{0}}%
\right) ^{2}\left( 1-\frac{3}{2}\frac{a}{U_{0}V_{0}}\right)\nonumber\right.\\
&\left.-\frac{5}{2}a%
\frac{\dot{U}_{0}\dot{V}_{0}}{U_{0}^{2}V_{0}^{2}}\right] ^{1/2}\tau \leq
\mathcal{L}_{\mathrm{ideal}}^{\text{GTD}}\left( \tau \right) \text{.}
\end{align}
These conditions resemble the above Eqs. \eqref{rese2} and extend the results, confirming the same functional behaviors.

\section{Efficiency from thermodynamic length}\label{sezione6}

In this section we are interested in understanding in concrete systems the difference between the formalisms of  thermodynamic geometry and GTD, highlighting the physical features of the above computed thermal lengths. To this end, we will introduce a novel concept of   \emph{efficiency}.

\subsection{Definition}

We propose a definition of efficiency expressed in terms of $r_{\mathcal{L}}
^{\text{GTD}}$ and $r_{\mathcal{L}}
^{\text{R}}$
\begin{subequations}
\begin{align}
&r_{\mathcal{L}}%
^{\text{GTD}}=%
\frac{\left\vert \mathcal{L}_{\mathrm{ideal}}^{\text{GTD}}\left(  \tau\right)
-\mathcal{L}_{\mathrm{vdW}}^{\text{GTD}}\left(  \tau\text{; }a\text{,
}b\right)  \right\vert }{\mathcal{L}_{\mathrm{vdW}}^{\text{GTD}}\left(
\tau\text{; }a\text{, }b\right)  },\label{relativediff1}\\
&r_{\mathcal{L}}%
^{\text{R}}=
\frac{\left\vert \mathcal{L}%
_{\mathrm{ideal}}^{\text{R}}\left(  \tau\text{; }R\right)
-\mathcal{L}_{\mathrm{vdW}}^{\text{R}}\left(  \tau\text{; }%
R\text{, }a\text{, }b\right)  \right\vert }{\mathcal{L}_{\mathrm{vdW}%
}^{\text{R}}\left(  \tau\text{; }R\text{, }a\text{, }b\right)
}\text{.}\label{relativediff2}%
\end{align}
\end{subequations}
Note in passing that $r_{\mathcal{L}}^{\text{R}}$ also depends on the constant $R$, beside $a,b$, whereas
$r^{\text{GTD}}_{\mathcal L}$ does not.

We refer to Appendices C and D for details on the calculations of $r_{\mathcal L}^{\text{GTD}}$ and $r_{\mathcal L}^{\text{R}}$ in Eqs. \eqref{relativediff1} and \eqref{relativediff2}, respectively.

Specifically, we can provide a new definition of efficiency
by resorting to the idea that in a geometric
formulation of mechanical evolution, one can quantify efficiency by the departure of an effective (non-geodesic evolution,
in general) from an ideal geodesic evolution (see also \cite{caf1}).
Such a geodesic evolution is characterized by paths of shortest
length that connect initial and final states.
We thus define a quantity that we henceforth call \textquotedblleft%
\emph{efficiency}\textquotedblright, as
\begin{equation}
\eta=\eta_{\text{Thermo}}^{\left(  \text{geometric}\right)  }\overset
{\text{def}}{=}
1-\frac{\left\vert \mathcal{L}_{\text{\textrm{real}}%
}-\mathcal{L}_{\mathrm{ideal}}\right\vert }{\mathcal{L}_{\text{\textrm{real}}%
}}\text{.}\label{efficiency}%
\end{equation}
We observe that $\eta$ in Eq. (\ref{efficiency}) is well defined for any real
gas and reaches its maximum value $1$ whenever the gas exhibits an ideal gas
behavior. Moreover, we note that $r_{\mathcal{L}}\overset{\text{def}}%
{=}\left\vert \mathcal{L}_{\text{\textrm{real}}}-\mathcal{L}_{\mathrm{ideal}%
}\right\vert /\mathcal{L}_{\text{\textrm{real}}}$ with $0\leq r_{\mathcal{L}%
}<1$ is the ideal-real relative change of the thermodynamic length with
respect to the reference length specified by the thermodynamic length of the
real gas. Clearly, when $r_{\mathcal{L}}$ approaches zero, the efficiency in
Eq. (\ref{efficiency}) approaches the maximum value $1$. Furthermore, the
minimum value of $\eta$ depends of the nature of the real gas. The more the
behavior of the real gas deviates from that of the ideal gas, the smaller the
value of $\eta$ is going to be and in any case, $0<\eta\leq1$. This latter condition, in turn, requires $|\mathcal L_{real}-\mathcal L_{ideal}|<\mathcal L_{real}$. Therefore, in order to satisfy this inequality in terms of thermodynamic lengths, we assume to work in the region of small deviations from ideality with $0<a,b\ll 1$. This working assumption is not physically unreasonable and is verified by a variety of common real gases characterized by van der Waals coefficients in the above mentioned ranges.

\subsection{Interpretation}

We here want to stress a fundamental property of $\mathcal L_{\rm real}$ and $\mathcal L_{\rm ideal}$. These two lengths are defined over two distinct manifolds whereas $s$ and $s_0$ lie on the same manifold. Thus, for the sake of notational simplicity, we denote with $U$ and $V$ the
thermodynamic parameters needed to parametrize a point on a manifold. Such
thermodynamic manifolds can describe distinct gases and, in addition, may be
equipped with distinct metrics. Therefore, despite our uniform notation, we
would like to clearly emphasize that we expect the various thermodynamical
parameters $\left\{  \left(  U\text{, }V\right)  \right\}  $ to specify
distinct behaviors of the trajectories on the manifolds depending on systems
and/or metrics being considered. Observe that in all the scenarios that we
consider, geodesic paths are specified by the nature of the gas and the chosen
metric on the manifold. Clearly, geodesic paths are specific to each scenario
and are not expected to be the same. However, for a given set of initial
conditions, we choose to observe the temporal evolution of our thermodynamic
parameters during an identical temporal duration equal to $\tau$. Moreover,
the freedom in the choice of the integration variables emerging from the
integration of geodesic equations allows us to suitably choose identical
initial conditions in all physical scenarios as well. Analogous examples come from classical physics, where the same commonly happens. For these reasons, we take the same notation throughout the work, albeit we know the two quantities lie on different manifolds. This has not modified in any way the overall treatment. To better underline this fact, we report an example in Appendix E.

Moreover, we point out that, in general, geometric measures of efficiency are thought as
being used to rank the performances of finite-time processes (for instance,
quantum driving processes or thermally driven processes) connecting two given
states (for instance, two quantum states or two thermodynamic states) on the
same manifold while optimizing a convenient figure of merit (for instance,
time, speed, available resources, entropy production). Within this framework
of thinking, the comparison occurs between an optimal (ideal, geodesic)
trajectory and an effective (real, non-geodesic) trajectory. Both trajectories
lie on the same manifold and connect the same pair of initial and final
states. In this scenario, the key idea is to distinguish geodesic from
non-geodesic paths and, in addition, efficient processes are the ones
connecting the two states by joining them along a trajectory that departs the
least from an ideal geodesic trajectory. Unlike these measures of efficiency,
our proposed measure aims at ranking processes by distinguishing a given
geodesic path (corresponding to a real gas, in our case) from a geodesic path
of reference (corresponding to an ideal gas, in our case). In addition, these
geodesic paths and the geodesic path of reference belong to distinct manifolds
and are generated under two constraints: Identical boundary conditions at the
starting time of observation and identical finite temporal duration for both
thermal processes being compared. Finally, efficient processes become the ones
yielding geodesic paths that depart the least from the geodesic path of
reference with the comparison being made by means of the relative difference
of the corresponding thermodynamic lengths.

\section{Comparison between Ruppeiner and GTD efficiencies}\label{sezione7}

In the case of Ruppeiner geometry and GTD,
from Eq. \eqref{efficiency}, we get

\begin{widetext}
\begin{align}
&\eta^{\text{R}}\left(  a\text{, }b\right)  =1-\frac{\left\vert \left(  \frac
{1-\frac{8}{5}a-\frac{6}{5}b-\frac{12}{5}ab^{2}+\frac{6}{5}a^{2}b-\frac{3}%
{5}a^{2}b^{2}+\frac{24}{5}ab-\frac{1}{5}a^{2}+\frac{3}{5}b^{2}}{\left(
a+1\right)  ^{2}\left(  b-1\right)  ^{2}}\right)  ^{1/2}-1\right\vert
}{\left(  \frac{1-\frac{8}{5}a-\frac{6}{5}b-\frac{12}{5}ab^{2}+\frac{6}%
{5}a^{2}b-\frac{3}{5}a^{2}b^{2}+\frac{24}{5}ab-\frac{1}{5}a^{2}+\frac{3}%
{5}b^{2}}{\left(  a+1\right)  ^{2}\left(  b-1\right)  ^{2}}\right)  ^{1/2}%
},\label{famoso1}\\
&\eta^{\text{GTD}}\left(  a\text{, }b\right)  =1-\frac{\left\vert \left(  \frac{4a+b+6ab^{2}%
-a^{2}b-12ab-2}{\left(  1-b\right)  \left(  1+a\right)  \left(
a-3ab-2\right)  }\right)  ^{1/2}-1\right\vert }{\left(  \frac{4a+b+6ab^{2}%
-a^{2}b-12ab-2}{\left(  1-b\right)  \left(  1+a\right)  \left(
a-3ab-2\right)  }\right)  ^{1/2}}\text{.}\label{famoso2}%
\end{align}
\end{widetext}

\noindent When $b=0$ and $a=0$, we have for
\begin{equation}
\eta^{\text{R}}\left(  a\right)  =1-\frac{\left\vert \left(
\frac{1-\frac{8}{5}a-\frac{1}{5}a^{2}}{\left(  a+1\right)  ^{2}}\right)
^{1/2}-1\right\vert }{\left(  \frac{1-\frac{8}{5}a-\frac{1}{5}a^{2}}{\left(
a+1\right)  ^{2}}\right)  ^{1/2}}\overset{a\ll1}{\approx}1-\frac{9}%
{5}a+\mathcal O\left(  a^{2}\right)  \text{,}%
\end{equation}
and,%
\begin{equation}
\eta^{\text{R}}\text{ }\left(  b\right)  =1-\frac{\left\vert \left(
\frac{1-\frac{6}{5}b+\frac{3}{5}b^{2}}{\left(  1-b\right)  ^{2}}\right)
^{1/2}-1\right\vert }{\left(  \frac{1-\frac{6}{5}b+\frac{3}{5}b^{2}}{\left(
1-b\right)  ^{2}}\right)  ^{1/2}}\overset{b\ll1}{\approx}1-\frac{2}%
{5}b+\mathcal O\left(  b^{2}\right)  \text{,}%
\end{equation}
respectively. For GTD, we have%

\begin{figure}\label{figura3}
\centering
\includegraphics[width=0.99\columnwidth,clip]{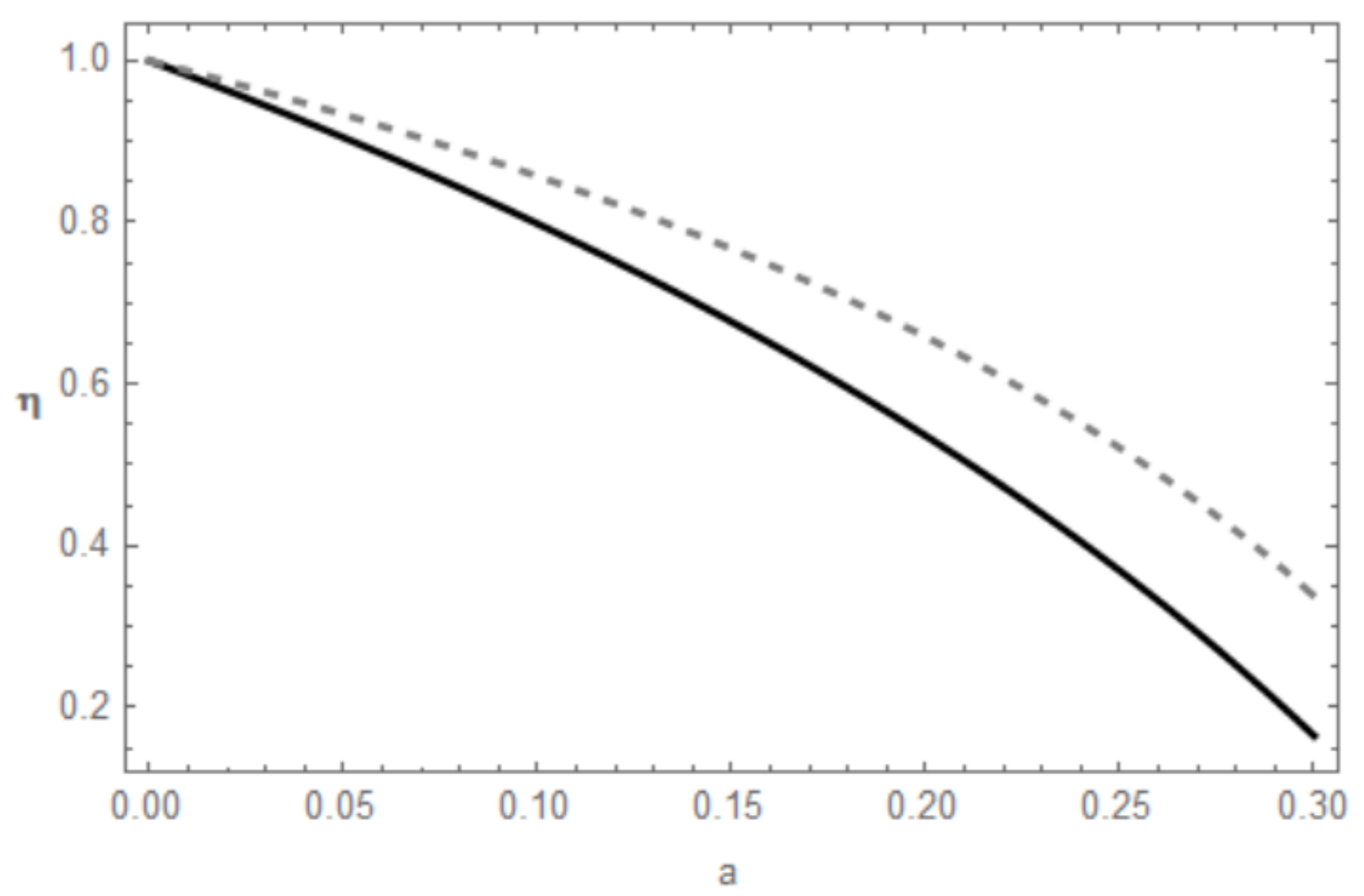}
\includegraphics[width=0.99\columnwidth,clip]{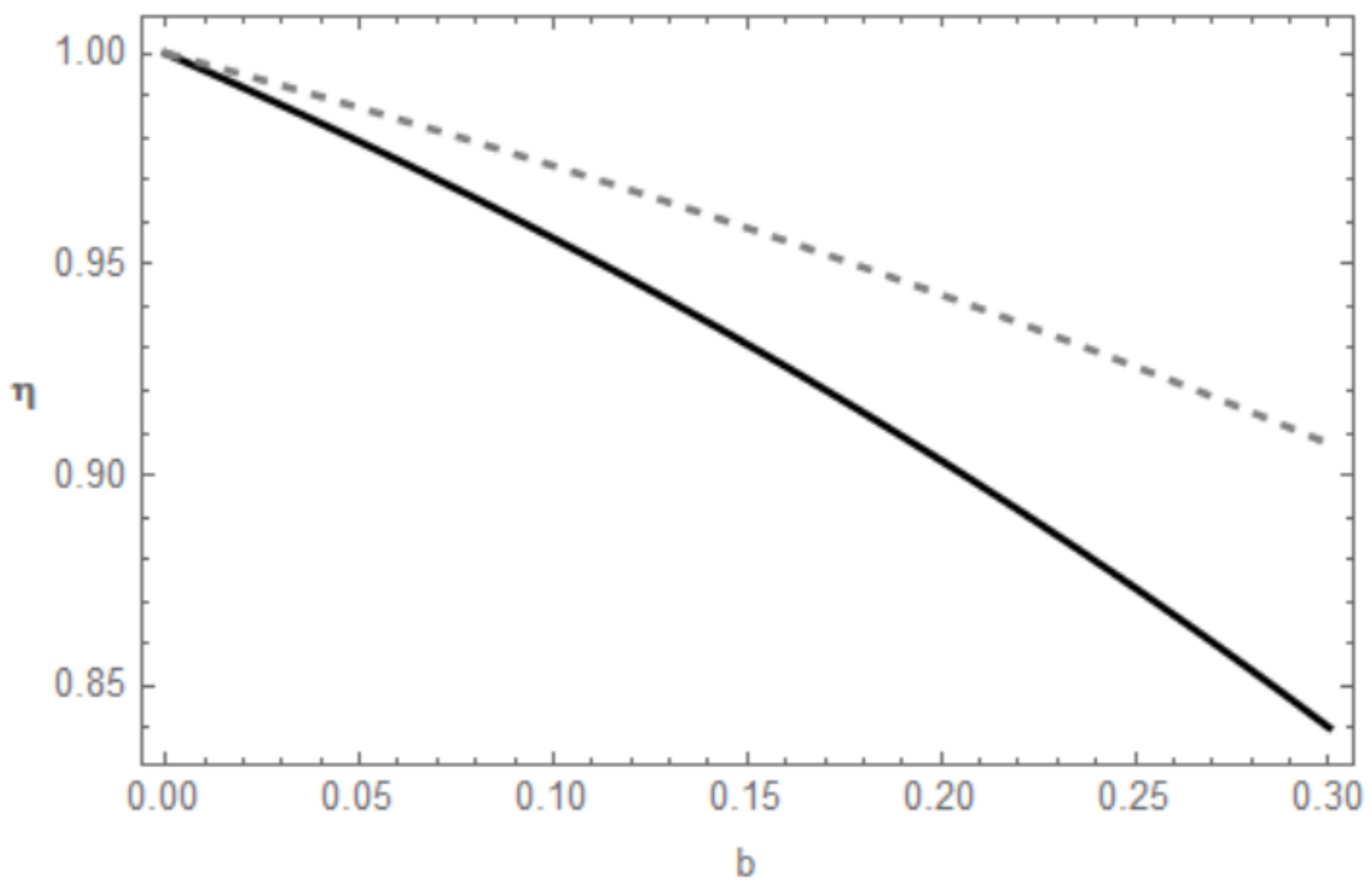}
\caption{\textit{Top panel}: plot of the geometric efficiency $\eta^{\text{R}}$ in Eq. \eqref{famoso1} as a
function of $a$ for $b$ fixed to $0$ (gray dashed line) and $0.3$ (black
dotted line). \textit{Bottom panel}: plot of the geometric efficiency $\eta^{\text{GTD}}$ in Eq.
\eqref{famoso2} as function of $b$ for $a$ fixed to $0$ (gray dashed line) and $0.3$
(black  line). As expected, the maximum of the efficiency
is reached at $(a,b)=(0,0)$ in both panels.}
\end{figure}

\begin{equation}
\eta^{\text{GTD}}\left(  a\right)  =1-\frac{\left\vert \left(  \frac
{4a-2}{\left(  1+a\right)  \left(  a-2\right)  }\right)  ^{1/2}-1\right\vert
}{\left(  \frac{4a-2}{\left(  1+a\right)  \left(  a-2\right)  }\right)
^{1/2}}\overset{a\ll1}{\approx}1-\frac{5}{4}a+\mathcal O\left(  a^{2}\right)  \text{,}%
\end{equation}
and,%
\begin{equation}
\eta^{\text{GTD}}\left(  b\right)  =1-\frac{\left\vert \left(  \frac
{2-b}{2\left(  1-b\right)  }\right)  ^{1/2}-1\right\vert }{\left(  \frac
{2-b}{2\left(  1-b\right)  }\right)  ^{1/2}}\overset{b\ll1}{\approx
}\allowbreak1-\frac{1}{4}b+\mathcal O\left(  b^{2}\right)  \text{,}%
\end{equation}
respectively. A direct comparison between the above approaches interestingly gives us a first result:
\begin{equation}\label{47}
\eta^{\text{R}}\left(  a\right)  \overset{a\ll1}{\approx}1-\frac{9}%
{5}a+\mathcal O\left(  a^{2}\right)  \leq1-\frac{5}{4}a+\mathcal O\left(  a^{2}\right),
\end{equation}
and,%
\begin{equation}\label{48}
\eta^{\text{R}}\text{ }\left(  b\right)  \overset{b\ll1}{\approx
}1-\frac{2}{5}b+\mathcal O\left(  b^{2}\right)  \leq1-\frac{1}{4}b+\mathcal O\left(
b^{2}\right),
\end{equation}
In both cases $\eta^{\text{R}}(a)\overset{a\ll1}{\approx}\eta^{\text{GTD}}\left(  a\right)$ and $\eta^{\text{R}}(b)\overset{b\ll1}{\approx}\eta^{\text{GTD}}\left(  b\right)$. The behaviors of $\eta^{\text{R}}$ and $\eta^{\text{GTD}}$ are reported in Fig. 3.

Again, the concept of efficiency is purely geometric and refers to the construction of our thermodynamic lengths, with no direct connection to standard thermodynamics.

We believe the inequalities in terms of thermal lengths in Eqs. \eqref{47} and \eqref{48} can be regarded as a signature of the fact that the complexity of the geodesic evolutions on manifolds equipped with the GTD metric might be lower than that on manifolds endowed with the Ruppeiner metric. Thus, it might appear more ``efficient'' distinguishing gases with the GTD metric. However, this remains a mere speculation at this point and demands further investigation. For a discussion on the conjectured links between the concepts of complexity and action
(or, equivalently, length), we refer to Refs. \cite{quaranta1,quaranta2}.

\section{Conclusions and outlook}\label{sezione8}

In this paper, we investigated the role played by the thermodynamic lengths for two intriguing cases of geometric thermodynamics. We first  evaluated  the  departures  from  ideal  to  real  gases  in a particular case of thermodynamic geometry and second we computed the same in the GTD formalism, where the Legendre invariance holds from the very beginning. We then formulated a strategy to relate thermodynamic lengths to a novel efficiency definition.
This originates primarily from the advantages it has:

\begin{itemize}
\item[{\bf 1.}] It provides a novel geometric way to distinguish ideal from
non-ideal thermal behaviors with all the benefits of an intuitive geometric
intuition of physical effects (see, e.g. \cite{alter0}).

\item[{\bf 2.}] It can help ranking deviations from \ ideality for a variety of
real gases.

\item[{\bf 3.}] It is a preliminary nontrivial step towards geometrically
quantifying the efficiency of actual (that is, effective, generally
non-geodesic) thermal evolutions in analogy to what can be achieved in quantum mechanics (see, e.g. \cite{caf1}).

\item[{\bf 4.}] It helps geometrically grasping the nature of different
metrizations on manifolds of equilibrium thermal states (see, e.g. \cite{alter1}).

\item[{\bf 5.}] It can be used to quantify the most suitable gas that characterizes particular frameworks, in which ideal gases break down and are unpredictive, e.g. cosmological applications, gravitational frameworks, and so forth (see, e.g. \cite{alter2,alter3}).

\end{itemize}

In addition to the aforementioned points,
to fully grasp the physical reasons of the presented efficiency measure, as well as its possible
limitations, we briefly compare the thermodynamic and geometric efficiency in what follows.
\begin{itemize}
\item In thermodynamics, the notion of efficiency of heat engines converting heat into useful work, for a cycle between two
reservoirs at temperature $T_{H}$ and $T_{C}$ (with $T_{H}>T_{C}$), is defined as the ratio of the performed work over the heat extracted from the high temperature reservoir \cite{adj3}. As a consequence of the second law of thermodynamics, it turns out that the efficiency
is upper bounded by the Carnot efficiency  $1-\frac{T_{C}}{T_{H}}$. A Carnot cycle is a reversible thermodynamic cycle with no total change of
entropy (reversibility condition) where the Carnot efficiency is obtained for
a quasistatic transformation which requires infinite time. Within this scheme, thermodynamic efficiency reflects \emph{the irreversibility
of the process and is applicable to expansion and compression processes where
heating or cooling takes place during the process}. In particular, efficient
thermodynamic processes are the ones that depart the least from the (ideal)
reversibility condition.

\item On the contrary,  a geometric definition of efficiency is quite different. In Ref. \cite{adj4}, thermodynamic geometry methods, together with the concept of
thermodynamic length, have been used to quantify the efficiency in thermal machines
close to equilibrium in both classical and quantum settings. In Ref. \cite{adj5}, methods of
information geometry combined with the concepts of thermodynamic length and
divergence of a path were employed to derive geometrical bounds of
irreversibility in both classical and quantum Markovian open systems. In Ref. \cite{adj6}, using thermodynamic geometry
arguments together with the notion of thermodynamic length, a universal
trade-off relation between efficiency and power in microscopic heat engines
driven (slow-driving regime) by arbitrary periodic temperature variations and
modulations of a mechanical control parameter was proposed.
\end{itemize}

Thus, our proposal is designed to take into account all the above arguments. Furthermore, our
definition of efficiency has also been motivated by the desire to make the most of the advantages
that a geometric description of thermodynamics implies. In particular, we point out the following
considerations.

It is known that the
efficiency of a Carnot cycle does not depend on the particular equations of
state that characterize the gas \cite{adj7}. In general, proofs of this universality can be
rather formal and one may naturally wonder if one can gain some physical
insight on this matter by using geometric formulations of thermodynamics and
show that the efficiency of a Carnot cycle is independent of the
\textquotedblleft\textit{working substance\textquotedblright}. It is also
well-known that friction has detrimental effects on the efficiency of heat
engines \cite{adj8}. Once again, one may
wish to gain some additional insights within the framework of geometric
thermodynamics on this latter matter by quantifying how the efficiency changes
when considering more realistic cycles with friction that \textquotedblleft%
\textit{deviate\textquotedblright} from a Carnot cycle in the original,
frictionless engine. Clearly, one also wonders on the available
\textquotedblleft\textit{choices of metrics\textquotedblright} (for example,
Ruppeiner and GTD\ metrics) on the thermal manifold needed to be employed.
More specifically, one may wonder why preferring one notion of
distinguishability over another. Motivated by these physical questions
concerning working substance independence, deviations from (ideal)
frictionless scenarios, and metric selections, we present in this paper the
first attempt (to the best of our knowledge) to introduce a geometric measure
of efficiency that aims at quantifying behavioral changes of a substance from
a reference substance in the framework of two distinct geometric
thermodynamics settings (specifically, the Ruppeiner and the GTD frameworks).

Furthermore, our choice could also be  non-unique and one may think of
proposing alternative \emph{indicators of deviations from ideality}. For example, it would be an interesting effort comparing such two distinct measures of
non-ideality by considering their suggested ranking on the same set of
different types of gases.


Using our proposed efficiency measure, we found the result that Legendre's invariance imposed \emph{a priori} in GTD seems to lead to a higher efficiency than that obtained in relativistic models in which this invariance is only imposed \emph{a posteriori}. Indeed, we observed that the evolution of a van der Waals gas on GTD manifolds seems to be more efficient than a van der Waals evolution on Ruppeiner's manifolds. As stated in Sect. VII, we have reason to believe this higher level of efficiency achieved within GTD can be viewed as a fingerprint of the fact that the complexity of geodesic paths on manifolds equipped with the GTD metric might be lower than that on manifolds endowed with the Ruppeiner metric. For this reason, it might appear more ``efficient'' quantifying deviations from ideality with the GTD metric. However, this statement is purely speculative at this stage, and we need to perform a detailed complexity-efficiency analysis to fully comprehend the physical meaning of our finding obtained here. 

We discussed the corresponding applications of our recipe in view of our efficiency and we explained the most relevant physical consequences of our approach.

It is worth remarking that the choice of $k=-1$ made within the GTD framework corresponds to a choice of coordinates, but according to the invariance of GTD it does not affect the quality of obtained results. We also leave the technical investigation on the
choice of more general and/or arbitrary boundary conditions together with the
possibility of ranking different types of substances with the help of thermal
lengths to future scientific efforts. As a conclusive remark on the choice of initial conditions, we reiterate that the
efficiency measure that we propose in this manuscript \emph{only depends on a
relative difference quantity constructed in terms of thermodynamic lengths}, namely  $\left\vert \mathcal{L}_{\text{real}}-\mathcal{L}_{\text{ideal}%
}\right\vert /$ $\mathcal{L}_{\text{real}}$.

Furthermore, since the GTD metric is Legendre invariant, we expect that also the GTD length and the GTD efficiency will be so (although a rigorous proof is needed).
Finally, the thermodynamic length turns out to be a measurable quantity, however the Ruppeiner length is measurable only within a specific potential, whereas the GTD length is measurable in all potentials.

For future works, we wish to address the problem of comparing the geodesic evolution of an arbitrary gas with its actual (that is, effective, generally non-geodesic) evolution. Moreover, we intend to deepen our understanding of our proposed geometric efficiency measure by extending, for instance, our investigation to large deviations from ideality for arbitrary multi-atomic real substances. We also want to investigate possible applications of our recipe in view of cosmological scenarios to select the correct gas driving the universe dynamics at given epochs.


\section*{Acknowledgements}
O.L. acknowledges the support of the Ministry of Education and Science of the Republic of Kazakhstan, Grant IRN AP08052311. This work was partially supported by UNAM-DGAPA- PAPIIT, Grant No. 114520, and Conacyt-Mexico, Grant No. A1-S-31269.

\newpage

\appendix

\section{GTD parameters}

Following the notation of Ref. \cite{adj10}, we
reported Eqs. \eqref{3a}, \eqref{3b}, and \eqref{gGTD} with the conformal factor $\Lambda=\Lambda\left(  Z^{A}\right)  $, representing
an arbitrary Legendre invariant function of the coordinates $Z^{A}=\left\{
\Phi\text{, }E^{a}\text{, }I^{a}\right\}  $. In what follows, we set
$\Lambda =\,\,\,$\textrm{constant }for the sake of computational simplicity.
Moreover, to deal with lengths of curves in an ordinary Riemannian geometric
setting, we choose $\Lambda$ to  assume a negative constant value equal to
$-1$ so that the GTD metric $g^{III}$ becomes a Riemannian metric exhibiting
both symmetry and positive-definiteness. For the sake of completeness, we
point out that it is possible to extend the concept of lengths of curves on
manifolds to a more general Riemannian framework such as a pseudo-Riemannian
setting. In such setting, the pseudo-Riemannian metric is not required to be
positive definite \cite{adj11}.

\section{Entropy of ideal and real gases}

Let us summarize some useful formulae for ideal and real gases, respectively, that we use throughout the text. For an ideal gas, the equation of state reads $PV=Nk_{B}T$, with $k_{B}$ being the Boltzmann constant, $N$
the number of molecules, and  $n\overset{\text{def}}{=}N/N_{A}$ is defined as
the number of moles of the gas with $N_{A}$ denoting Avogadro's number. Its  internal energy is a function of the temperature only and reads $U_{\mathrm{ideal}}(T)=nC_{V}T$. Using the first law of thermodynamics, $dQ=T dS =dU+PdV$, we immediately get $
dS=nC_{V}\frac{dU}{U}+nR\frac{dV}{V}$, or equivalently%
\begin{equation}
S_{\mathrm{ideal}}\left(  U\text{, }V\right)  =nC_{V}\log\left(  U\right)
+nR\log\left(  V\right)  +\mathcal{C}_i, \label{entropyideal}%
\end{equation}
with $\mathcal{C}_i$ an integration constant. In the van der Waals gas, the equation of state reads $
\left(  P+\frac{n^{2}}{V^{2}}a\right)  \left(  V-nb\right)  =nRT$ and the  internal energy reads
$U_{\mathrm{VdW}}(T)=nC_{V}T-n^{2}\frac{a}{V}+U_0$. We thus have $
dS=\frac{nC_{V}}{U+n^{2}\frac{a}{V}}dU+\frac{nR}{V-nb}dV$ or equivalently
\begin{equation}
S_{\mathrm{vdW}}\left(  U\text{, }V\right)  =nC_{V}\log\left(  U+n^{2}\frac
{a}{V}\right)  +nR\log\left(  V-nb\right)  +\mathcal{C}_r,
\label{entropyreal}%
\end{equation}
with $\mathcal{C}_r$ an integration constant.

\section{Computation of thermodynamic metrics}

Recalling that the Ruppeiner metric is defined as $
g=g_{ab} \left(  \theta\right)  d\theta^{a}\otimes d\theta^{b}$,
where, $
g_{ab}\left(  \theta\right)  \overset{\text{def}}{=}-\frac
{\partial^{2}S\left(  \theta\right)  }{\partial\theta^{a}\partial
\theta^{b}}$,
with $\theta=\left(  \theta^{1}\text{, }\theta^{2}\right)  =\left(  U\text{,
}V\right)  $, we have for the ideal and real gases

\begin{align}
g_{11}^{ideal}\left(  U\text{, }V\right)   &  =\frac{nC_{V}}{U^{2}}\text{,
}\nonumber\\
g_{12}^{ideal}\left(  U\text{, }V\right)   &  =g_{21}\left(  U\text{, }V\right)
=0\text{,}\nonumber\\
g_{22}^{ideal}\left(  U\text{, }V\right)   &  =\frac{nR}{V^{2}}\text{.}\\
g_{11}^{vdW}\left(  U\text{, }V\right)   &  =nC_{V}\frac{V^{2}}{\left(
an^{2}+UV\right)  ^{2}}\text{, }\nonumber\\
g_{12}^{vdW}\left(  U\text{, }V\right)   &  =g_{21}\left(  U\text{, }V\right)
=-\frac{an^{3}C_{V}}{\left(  an^{2}+UV\right)  ^{2}}\text{,}\nonumber\\
g_{22}^{vdW}\left(  U\text{, }V\right)   &  =\frac{nR}{V^{2}}\frac{V^{2}}{\left(
V-bn\right)  ^{2}}-\frac{an^{3}C_{V}}{V^{2}}\frac{\left(  an^{2}+2UV\right)
}{\left(  an^{2}+UV\right)  ^{2}}\text{.}\nonumber
\end{align}

\noindent In the case of GTD, we use the most general metric
\be
	g =  \Lambda\left(E^a \frac{\partial \Phi}{\partial E^a}
	\right)^{2k+1} \frac{\partial^2 \Phi}{\partial E^a \partial E^b}
	\, \d E^a \otimes \d E^b \,.
\ee
and obtain, with $\Lambda=-1$
\begin{widetext}
\begin{align}
g_{11}^{ideal}\left(  U\text{, }V\right)    & =\frac{(nC_{V})^{2k+2}}{U^{2}%
}\text{,}\\
& \\
g_{12}^{ideal}\left(  U\text{, }V\right)    & =g_{21}^{ideal}\left(  U\text{,
}V\right)  =0\text{,}\\
& \\
g_{2}^{ideal}\left(  U\text{, }V\right)    & =\frac{\left(  nR\right)
^{2k+2}}{V^{2}}\text{,}\\%
g_{11}^{vdW}\left(  U\text{, }V\right)    & =\left(  nC_{V}\frac{U}%
{U+n^{2}\frac{a}{V}}\right)  ^{2k+1}\frac{nC_{V}V^{2}}{\left(  an^{2}%
+UV\right)  ^{2}}\text{,}\\
& \\
g_{12}^{vdW}\left(  U\text{, }V\right)    & =g_{21}^{vdW}\left(  U\text{,
}V\right)  =-\frac{1}{2}\left[  \left(  nC_{V}\frac{U}{U+n^{2}\frac{a}{V}%
}\right)  ^{2k+1}+\left(  nR\frac{V}{V-nb}-nC_{V}\frac{n^{2}\frac{a}{V}%
}{U+n^{2}\frac{a}{V}}\right)  ^{2k+1}\right]  \frac{an^{3}C_{V}}{\left(
an^{2}+UV\right)  ^{2}}\text{,}\\
& \\
g_{22}^{vdW}\left(  U\text{, }V\right)    & =-\left(  nR\frac{V}{V-nb}%
-nC_{V}\frac{n^{2}\frac{a}{V}}{U+n^{2}\frac{a}{V}}\right)  ^{2k+1}\left[
-\frac{nR}{V^{2}}\frac{V^{2}}{\left(  V-bn\right)  ^{2}}+\frac{an^{3}C_{V}%
}{V^{2}}\frac{\left(  an^{2}+2UV\right)  }{\left(  an^{2}+UV\right)  ^{2}%
}\right]  \text{.}%
\end{align}
\end{widetext}

\section{Computation of thermodynamic lengths}

In what follows, we set $n=1$, $k=-1$, $C_{V}=\left( 3/2\right) R$
(monoatomic gas), and assume boundary conditions given by $U_{0}\overset{%
\text{def}}{=}U(0)=1$, $V_{0}=V\left( 0\right) =1$, $\dot{U}_{0}\overset{%
\text{def}}{=}\left( dU/d\xi \right) _{\xi =0}=1$, and $\dot{V}_{0}=\left(
dV/d\xi \right) _{\xi =0}=1$. Then, $\mathcal{L}_{\mathrm{ideal}}^{\text{%
\textrm{R}}}$ in Eq. \eqref{lideal1}, $\mathcal{L}_{\mathrm{vdW}}^{\text{\textrm{R}}}$
in Eq. \eqref{10}, $\mathcal{L}_{\mathrm{ideal}}^{\text{\textrm{GTD}}}$ in Eq.
\eqref{20}, and $\mathcal{L}_{\mathrm{vdW}}^{\text{\textrm{GTD}}}$ in Eq. \eqref{conlu}
become,%
\begin{widetext}
\begin{align}
\mathcal{L}_{\mathrm{ideal}}^{\text{R}}\left(  \tau\right)   &
=\left(  \frac{5}{2}R\right)  ^{1/2}\tau\text{,}\nonumber\\
\mathcal{L}_{\mathrm{ideal}}^{\text{GTD}}\left(  \tau\right)   &  =2^{1/2}%
\tau\text{,}\nonumber\\
\mathcal{L}_{\mathrm{vdW}}^{\text{R}}\left(  \tau\right)   &  =\left[
\frac{1-\frac{8}{5}a-\frac{6}{5}b-\frac{12}{5}ab^{2}+\frac{6}{5}a^{2}%
b-\frac{3}{5}a^{2}b^{2}+\frac{24}{5}ab-\frac{1}{5}a^{2}+\frac{3}{5}b^{2}%
}{\left(  a+1\right)  ^{2}\left(  b-1\right)  ^{2}}\right]  ^{1/2}\left(
\frac{5}{2}R\right)  ^{1/2}\tau\text{,}\nonumber\\
\mathcal{L}_{\mathrm{vdW}}^{\text{GTD}}\left(  \tau\right)   &  =\left[
\frac{4a+b+6ab^{2}-a^{2}b-12ab-2}{\left(  1-b\right)  \left(  1+a\right)
\left(  a-3ab-2\right)  }\right]  ^{1/2}2^{1/2}\tau\text{,} \label{FU2}%
\end{align}
\end{widetext}
respectively, where the superscript ``$\text{R}$'' refers to the word \emph{Ruppeiner}.

This superscript should not be confused with the gas constant, ``$R$''.

\section{Freedom in choosing the  integration constants: a physical example}

For the sake of completeness, we report here a physical example on the freedom in the choice of the integration constants. In particular, it
emerges from the integration of geodesic equations and allows us to suitably choose identical initial conditions in
all physical scenarios as well. To clarify this point, consider the one-dimensional motion of a falling body in the presence of uniform gravity satisfying the equation
$\dot v_1=g$ with $v_1$  the velocity and $g$ the acceleration of gravity. Moreover, consider the one-dimensional
motion of a falling body in the presence of uniform gravity and air resistance satisfying the relation $\dot v_2 = g-\frac{b}{m}v_2$ with $b$  the air resistance coefficient. In the former scenario, we have $v_1(t)=A_1+gt$, $\forall A_1\in \mathbb{R}$. In the latter scenario, instead, we get $v_2(t)=\frac{mg}{b}-\frac{A_2}{b}e^{-\frac{b}{m}t}$, $\forall A_2\in \mathbb{R}$.

Clearly, the two velocity trajectories $v_1(t)$
and $v_2(t)$ are distinct. However, thanks to the freedom in choosing the integration constants $A_1$ and $A_2$, one
can impose $v_1(0) = v_0 = v_2(0)$ with the consequence that $A_1 = v_0$ and $A_2 = mg - bv_0$. Therefore, one can
choose identical initial conditions and the consequence of this choice will be encoded in the functional form of
the integration constants.

\end{document}